\documentclass[12pt]{article}
\usepackage[dvips]{graphics,color,epsfig}
\usepackage{amsmath,amssymb}
\usepackage[square,comma,sort&compress]{natbib}
\renewcommand{\Im}{\hbox{Im\,}}
\renewcommand{\Re}{\hbox{Re\,}}

\title{\vspace{-10mm}\bf \small Theory of four-wave mixing and accompanying dissociation and population transfer
controlled with laser-induced continuum structures}\vspace{-15mm}
\author{\small $^1$V. V. Kimberg,
$^{1,2}$A.K. Popov\thanks{e-mail: popov@iph.krasn.ru;
http://www.kirensky.ru/popov}, and $^2$Thomas F.
George\thanks{e-mail: tgeorge@uwsp.edu; http://www.uwsp.edu/admin/chancellor/tgeorge}\\
\small $^1$L.V. Kirenskii Institute of Physics, Siberian Division,\\
\small Russian
Academy of Sciences, Krasnoyarsk, 660036, Russia\\
\small $^{2}$Office of the Chancellor/ Departments of Chemistry\\
\small and
Physics \& Astronomy,\\ \small University of Wisconsin-Stevens Point,\\
\small Stevens Point, WI 54481-3897, USA\\
\\
\small A chapter in {\it Recent Research Developments in Optics},
vol. 2,\\ \small Research Signpost, Trivandrum, India, 2002\\
\\
Dated: July 29, 2002}
\date{}
\begin{document} \maketitle
\begin{abstract}
A theory is developed and applied to the study of opportunities
and specific features  of coherent control of four-wave mixing as
well as of the accompanying processes in the continuous-wave
regime, which involve transitions between bound and free quantum
states. Such opportunities become feasible through constructive
and destructive interference of quantum pathways. Two coupling
schemes of practical importance are investigated. In the first, a
ladder energy level scheme, fully-resonant sum-frequency
nonlinear-optical generation of short-wavelength radiation driven
by several strong fields is investigated. The relaxation processes
as well as absorption of the fundamental and generated radiations,
which play an important role, are taken into consideration.   It
is shown that the generation output can be considerably increased
through the appropriate adjustment of several laser-induced
continuum structures.  In the second, a folded scheme, a possible
control of two-photon dissociation ($\Lambda$-scheme) using
auxiliary laser radiation applied to the adjacent bound-free
transition ($V$-configuration) is investigated. Besides
dissociation, the proposed method  enables one to control
population transfer between two upper discreet levels via the
lower-energy dissociation continuum, while direct transition
between these states is not allowed. The opportunities of
manipulating these processes as well as of four-wave-mixing-based
spectroscopy are explored both analytically and through a
numerical simulation  for Na$_2$ dimers.

 PACS: 42.50.Hz,
42.50.Ct,42.65.Ky, 42.65.Yj, 32.80.Fb., 32.80.Qk
\end{abstract}

\thispagestyle{empty}\newpage %\tableofcontents
\section*{Contents}
{\bf 1. Introduction}\hfill 2 \\

\noindent{\bf 2. Four-wave mixing, absorption and refractive
indices in ladder \\ \hspace*{5mm} energy-level schemes controlled
by interference of two
LICS}\hfill 4\\
\hspace*{5mm} 2.1. Equations for coupled electromagnetic waves and
their solution\hfill 4 \\
\hspace*{5mm} 2.2. Density matrix master equations and their
solutions\hfill 5\\
\hspace*{5mm} 2.3. Laser-induced structures in absorption and
refraction spectra\\ \hspace*{15mm} at discrete and continuous
transitions\hfill
9\\
\hspace*{5mm} 2.4. Resonance sum-frequency generation in
strongly-absorbing media \\ \hspace*{15mm} enhanced by quantum
interference\hfill 12\\
\hspace*{5mm} 2.5. Absorption and dispersive spectra at
Doppler-broadened transitions\hfill 17 \\

\noindent{\bf 3. Four-wave mixing, dissociation and population
transfer controlled \\ \hspace*{5mm} by  LICS in folded energy-level schemes}\hfill 19\\
 \hspace*{5mm} 3.1. Quasistationary solution of
density matrix equations \hfill 21\\ \hspace*{15mm} 3.1.1. Open
configuration \hfill 24 \\ \hspace*{15mm} 3.1.2. Closed
configuration\hfill 25\\ \hspace*{5mm} 3.2.
Numerical analysis\hfill 26 \\
\hspace*{15mm} 3.2.1. Coherent control of populations and
dissociation  in
folded schemes\hfill 27 \\
\hspace*{15mm} 3.2.2. Coherent control of generated radiation in
folded schemes\hfill 29\\

\noindent {\bf Conclusions}\hfill 33\\

\section{Introduction}
The opportunities of laser control of optical properties and
chemical reactivity of substances associated with the continuum
energy states of atoms and molecules, and based on nonlinear
quantum coherence and interference processes, have attracted
considerable attention since their theoretical prediction
\cite{GP2,GP3} and experimental realization \cite{GP9}. These
effects were shown to provide  the feasibility of manipulating the
spectral characteristics of the absorption, refraction, and
nonlinear-optical conversion of weak radiations by an additional
strong field, which is in resonance with a transition between an
excited vacant level and a state in the continuum. The
strong-field effects in such coupling configuration  manifest
themselves by the appearance of spectral autoionizing-like
laser-induced continuum structures (LICS). In the presence of a
real resonance autoionizing level a strong field  has a strong
influence on the spectral characteristics of this resonance too
\cite{AI1,AI2,AI3,Gel}. The effects of the LICS on the absorption,
refraction, nonlinear-optical generation and photoionization may
differ significantly (see, for example, Refs
{\cite{Gel,Kn,Fau,2a,2c,Cav,2d1,bull,Kimb,2e}} and the references
therein). Most of the recent publications consider LICS in
photoionization. The influence of real autoionizing levels and of
the LICS  on nonlinear susceptibility and four-wave mixing (FWM)
has been investigated  for frequencies of the other fields far
from one-photon resonances \cite{2c,Kimb}. The interference of two
LICS was shown to bring new qualitative effects in spectral
properties of absorption, refraction and nonlinear optical
generation of short-wavelength radiation \cite{Kimb}.

Recently, the laser and chemical communities have focused on
coherent quantum control of chemical reactions and other
dissipation processes like photodissociation of molecules and
photoionization of atoms
\cite{TR1,BrSh,TR2,Br1,D1,Sh1,D2,U1,U2,1D,1DD,D4,D3,
Br2,R1,Br3,Br4,Br5,D5,D8,D9,R2,R3,R4,Gor1,Sh2,Sh3,
102,Mcool,R5,R6,R9,Band,Br6,R7,R8,117}. In many cases, such
control is based on implementation of the interference of
different quantum channels, which give rise to laser-induced
continuum structures (LICS) in the dissociation and ionization
continua. Recent experiments aimed at control of photoionization
of metastable helium atoms and some other processes making use of
this technique are reported elsewhere \cite{Half, Yats, Berg}.

This paper further develops theoretical approaches to these
problems and reports novel opportunities for manipulating spectral
characteristic of short-wavelength generation and
photodissociation through the interplay of two LICS induced by
strong continuous-wave laser fields coupling the dissociation
continuum to two different bound states. Primary attention is paid
to two important applications: (1) coherent control of
sum-frequency short-wavelength generation in a ladder-type
energy-level scheme (depicted in Fig. \ref{leva}) and (2) coherent
control of dissociation, population transfer and four-wave mixing
feasible in a folded-type scheme (depicted in Fig. \ref{levb}).

As concerns the first problem,  we generalize our investigation of
the LICS to the case of several strong control fields  in
situations when one can expect the strongest enhancement of the
generated power. We consider the combined influence of the
interference of two laser-induced structures in the continuum and
of the laser-induced transparency (for the one-photon-resonant
initial radiation and for the generated radiation) on the
processes of nonlinear-optical generation of short-wavelength
radiation. We demonstrate for the first time that nonlinear
interference resonances in the output power of the  radiation
generated in an absorbing medium may differ considerably from the
corresponding resonances in a transparent medium because of the
combined influence of the nonlinear resonances in the nonlinear
polarization and in absorption of both the initial and generated
radiations.

A two-photon dissociation ($\Lambda$-scheme) controlled by an
auxiliary laser field coupling continuum with another discrete
state ($V$-configuration) is investigated for the second problem.
All the coupled radiations are strong enough to drive molecular
transitions. The proposed technique enables the coherent laser
control of dissociation related to the lower laying molecular
'dark' states that are not connected with the ground one by the
allowed transition.  On the other hand, the scheme under
consideration enables one to transfer the population between two
upper bound states which are not connected directly by the allowed
transition. An analytical solution of the corresponding master
density matrix equations of the problem is obtained, and a
numerical analysis for relevant experimental conditions
\cite{{Half,Berg}} is performed. The possibility of manipulating
the dissociation spectra and populations of excited states at the
expense of the interference of quantum pathways through a variety
of continuum states is explored. The dependence of the effects on
the composed Fano parameters related to high excited levels, on
the detuning between two LICS and on the intensities of the laser
radiation is investigated.
\section{Four-wave mixing, absorption and refractive
indices in ladder energy-level schemes
controlled by interference of two LICS}
\subsection{Equations for coupled electromagnetic waves and their solution}
Let us consider four plane-polarized electromagnetic waves
travelling along the $z$ axis of an isotropic medium,
\begin{align}
E^{j}(z,t)={\rm Re}\{E_{j}(z)\exp [{\rm i}(\omega _{j}t-k_{j}z)]\},
\end{align}
where $k_{j}$ is the complex wave-number corresponding to the
frequency $\omega _{j}$ ($j=1,2,3,S $). We assume that the fields
$E_{1}$ and $E_{S }$ are weak compared to the driving fields
$E_{2}$ and $E_{3}$, which do not vary along the medium. On the
contrary, the fields $E_{1}$ and $E_{S}$ can change considerably
along the medium,  because of both absorption and
nonlinear-optical conversion. Then the spacial behavior of the
waves $E_{S}$ and $E_{1}$ is  described by two coupled reduced
wave equations,
\begin{align}
dE_{S }(z)/dz&={\rm i}2\pi k_{S}'\chi ^{(3)}_{S}E_{2}E_{3}E_{1}(z)\exp ({\rm
i}\Delta kz),
\nonumber\\
dE_{1}(z)/dz&={\rm i}2\pi k_{1}'\chi ^{(3)}_{1}
E^{*}_{2}E^{*}_{3}E_{S }(z)\exp (-{\rm i}\Delta
kz)\label{volsisl}.
\end{align}
Here $k_j = k_j'-{\rm i}k_j''=(2\pi\omega_j/c) \chi_j$,
$k_j''=\alpha _{j}/2$, and $ \chi_j, \alpha _{j}$ are the
effective linear susceptibilities and absorption indices for the
corresponding radiations, and $\chi ^{(3)}_{1}, \chi ^{(3)}_{S }$
are the nonlinear susceptibilities describing the four-wave mixing
processes: $\omega _{S }\leftrightarrow \omega _{1}+\omega
_{2}+\omega _{3}$, $\Delta k=k_{S }-k_{1}-k_{2}-k_{3}$. The
quantum conversion efficiency of the radiation $E_1$ into $E_{S}$
varies along the medium as
\begin{align}
\eta_{{\rm q}}=(k_1'/k_S')|E_S(z)/E_1(0)|^2\exp(-\alpha_Sz)\label{qu}.
\end{align}

Let us first  consider the case of low efficiency, for which the
change in the $E_1$ caused by the nonlinear-optical conversion can
be ignored. Then the second equation in (\ref{volsisl}) can be
ignored as well and, with the account of the boundary condition
$E_S (z=0)=0$, one obtains the following for the generated
radiation and quantum conversion efficiency:
\begin{align}
&E_S(z)=(2\pi k_S'/ \Delta k)\chi_S^{(3)} E_1 E_2 E_3 [\exp (-{\rm
i}\Delta k z)-1],\\
&\eta_{\rm q}(z)= {k_S'}{k_1}'(\big|2\pi\chi_S^{(3)} E_2
E_3\big|^2/|\Delta k|^2)\exp(-\alpha_S z) \big|\exp(-{\rm i}\Delta
k z)-1\big|^2.\label{qus}
\end{align}
If the medium length is much shorter than the minimum absorption
length $L_{abs}=\min\{L_1=2/\alpha_1, L_S=2/\alpha_S\}$ and both
of these are assumed much shorter than the coherence length
$L_{coh}=\Delta k'^{-1}$, then Eq. (\ref{qus}) reduces to the
approximate formula
\begin{align}
\eta_{{\rm q}}={k_S}' {k_1}'\big|2\pi\chi_S^{(3)} E_2 E_3\big|^2
L_{\rm e}^2, \label{kep}
\end{align}
where $L_{\rm e}$ represents either the length of the medium (in
the case of weak absorption) or the optimal length of the order of
$\min\{L_{abs}, L_{coh}\}$.

For a medium with substantial absorption dispersion
($\alpha_1\neq\alpha_S$), but $\Delta k'=0$ and $\chi
^{(3)}_{1}=\chi ^{(3)*}_{S }$, the solution of equations
(\ref{volsisl}), for the more general case of considerable
conversion, takes the form \cite{tim}
\begin{align}
\eta_{\rm q}(z)=4\frac{\tilde\eta_{{\rm q}0}}{|b|}\exp
\left[-(\alpha_1+\alpha_{S})\frac{z}{2}\right] \left[{\rm
sinh}^{2}\left(\sqrt{\frac{(|b|-b)}{2}}\frac{z}{2}\right)+\sin ^{2}
\left(\sqrt{\frac{(\mid b\mid+b)}{2}}\frac{z}{2}\right)\right].\label{qe}
\end{align}
Here $\tilde{\eta }_{q0}=k_{1}'k_{S }'\mid 2\pi \chi ^{(3)}_{S
}E_{2}E_{3}\mid ^{2}$ is the conversion efficiency per unit of the
medium length under constant fundamental radiations, and
$b=4\tilde{\eta }_{q0}-(\alpha _{1}-\alpha _{S })^{2}/4$ defines
the difference between the rates of absorption and
nonlinear-optical conversion of the radiations. If $b<0$, the
conversion rate of  $\hbar\omega _{1}$ photons into  $\hbar\omega
_{S }$ photons is less than their absorption rate; if $b=0$, the
photon absorption and conversion rates are equal; and if $b>0$,
the nonlinear-optical conversion rate exceeds the absorption rate.
In the latter case, one can expect an oscillatory dependence of
the transfer of the weak radiations from one to the another and
back along the medium.

The susceptibilities $\chi ^{(3)}_{S }$, $\chi^{(3)}_{1}$,
$\chi_{S }$ and $\chi_{1}$ can be derived from the medium
polarizations at the corresponding frequencies:
\begin{align}
P(\omega _{j })=\chi_{j }E_{j},\quad P^{\rm NL}(\omega _{S })=\chi
^{(3)}_{S }E_{1}E_{2}E_{3},\quad P^{\rm NL}(\omega _{1})=\chi
^{(3)}_{1}E_{S }E^{*}_{2}E^{*}_{3}\label{pnl}.
\end{align}
These components can be calculated conveniently with the aid of a
density matrix, $\rho _{ij}$, as
\begin{equation}
 \mathcal P(\omega _{j})=N\rho _{ij}(\omega _{j}) d_{ji}+ {\rm c.c.}, \label{10}
\end{equation}
where $N$ is the atomic number density in the medium, and ${
d}_{ji}$ is a matrix element of the projection of the transition
electric-dipole moment along the direction of the electric vector
of the corresponding field. The problem of finding and optimizing
the quantum efficiency of the conversion process thus reduces to
finding the off-diagonal elements of the density matrix.
\subsection{Density matrix master equations and their solutions}
\begin{figure}[!h]
%{(a)}\hspace{80mm} {(b)}
\begin{center}
\includegraphics[width=0.3\textwidth]{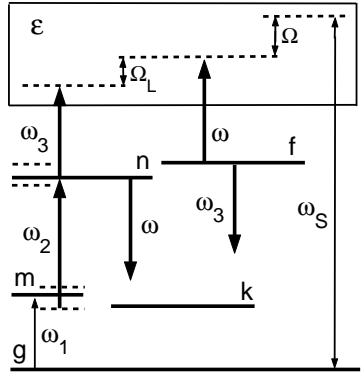}
\end{center}
\caption{\label{leva} LICS-based coherent control in ladder
schemes.}
\end{figure}

First, we shall consider the transition scheme depicted  in Fig.
\ref{leva}. A strong field ${\mathit E}_{2}$ at frequency
$\omega_{2}$ is close to  resonance with the transition between
 levels $m$ and $n$, while the strong fields ${\mathit E}$ and
${\mathit E}_{3}$ at frequencies $\omega $ and $\omega_{3}$ are in
resonance with the transitions between levels $f$ and $n$ and some
states in the continuum. Radiation at the frequency $\omega_S$ can
be either a probe or generated by four-wave mixing. We shall
investigate the influence of the strong fields on the spectral
characteristics of the absorption of the radiations ${\mathit
E}_{1}$ and ${\mathit E}_{S}$ at the frequencies $\omega_{1}$ and
$\omega_S$ (both as independent probe radiations or,
alternatively, as the frequencies linked through generation,
$\omega_S=\omega_1+\omega_2+\omega_3$). The field $E_1$ is assumed
close to resonance with the transition from the ground state to
the level $m$ and $E_S$ to the transition from the ground state to
the continuum. These radiations are weak, so that a change in the
level populations due to all resonant couplings can be ignored. We
neglect the degeneration of all states, including those in the
continuum. In addition to the contribution of the off-resonant
continuum states, we account for the contribution of discrete
off-resonant levels which are combined to form the level $k$ in
Fig. \ref{leva}. We suppose that the detunings $\mid\omega
_{1}$-$\omega _{gm}\mid$, $\mid\omega _{1}+\omega _{2}-\omega
_{gn}\mid$ and $\mid \omega -\omega _{3}-\omega _{nf}\mid$ are
considerably less than all the other detunings.
The equations for the density matrix, considered in the
interaction representation to within the first order of
perturbation theory in weak fields, can be written as follows:
\begin{align}
&d\rho _{gm}/dt+\Gamma _{gm}\rho _{gm}={\rm i}(\rho
_{gg}V_{gm}+\rho _{gn}V_{nm}),
\nonumber\\
&d\rho_{gn}/dt+\Gamma _{gn}\rho _{gn}={\rm i}\int \rho
_{g\varepsilon }V_{\varepsilon n}d\varepsilon +{\rm i}\rho
_{gm}V_{mn}+{\rm i}\sum \rho _{gk}V_{kn},
\nonumber\\
&d\rho _{gf}/dt+\Gamma _{gf}\rho _{gf}={\rm i}\int \rho
_{g\varepsilon }
V_{\varepsilon f}d\varepsilon +{\rm i}\sum \rho _{gk}V_{kf}\label{mp},\\
&d\rho _{g\varepsilon }/dt={\rm i}(\rho _{gg}V_{g\varepsilon
}\hbox{{+}}\rho _{gn}V_{n\varepsilon }\hbox{{+}}\rho
_{gf}V_{f\varepsilon }),
\nonumber\\
&d\rho _{gk}/dt+\Gamma _{gk}\rho _{gk}={\rm i}(\rho
_{gn}V_{nk}+\rho _{gf}V_{fk}). \nonumber
\end{align}
\noindent Here, the index $\varepsilon$ denotes the continuum
states; $V_{mn}=G_{mn}\exp [{\rm i}(\omega _{2}-\omega _{nm})t],
V_{g\varepsilon }=G_{g\varepsilon }\exp [{\rm i}(\omega _{S
}-\omega _{g\varepsilon })t], V_{n\varepsilon }=G_{n\varepsilon
}\exp [{\rm i}(\omega _{3}-\omega _{n\varepsilon })t],
V_{gm}=G_{gm}\exp [{\rm i}(\omega _{1}-\omega _{gm})t],
V_{f\varepsilon }=G_{f\varepsilon }\exp [{\rm i}(\omega -\omega
_{f\varepsilon })t], V_{kn}=G_{kn}\exp [{\rm i}(\omega -\omega
_{kn})t], V_{kf}=G_{kf}\exp [{\rm i}(\omega _{3}-\omega _{kf})t]$
are the matrix elements of the Hermitian interaction Hamiltoman
$\hat V$, considered in the electric-dipole approximation and in
the interaction representation (in units of $\hbar$); $V_{ij}$ =
$V^{*}_{ji}; G_{mn}=-{ E}_{2}{ d}_{mn}/2\hbar , G_{g\varepsilon
}=-{ E}_{4 }{ d}_{g\varepsilon }/2\hbar,$ etc.; and $\Gamma _{ij}$
is the homogeneous half-width of the $i-j$ transition. In the
approximation of the weak fields $E_1$ and $E_S$, we obtain $\rho
_{gg}\approx 1,\rho _{m}\approx \rho _{n}\approx \rho _{f}<<1.$
\par
Under steady-state conditions, each off-diagonal element of the
density matrix is a sum of two components, which  may oscillate at
different frequencies:
\begin{align}
\rho _{g\varepsilon }&=r_{g\varepsilon }\exp [{\rm i}(\omega _S-\omega
_{g\varepsilon })t]+R_{g\varepsilon }\exp [{\rm i}(\omega _{1}+\omega
_{2}+\omega _{3}-\omega _{g\varepsilon })t],
\nonumber\\
\rho _{gn}&=r_{gn}\exp [{\rm i}(\omega _S-\omega _{3}-\omega
_{gn})t]+R_{gn}\exp [{\rm i}(\omega _{1}+\omega _{2}-\omega _{gn})t],
\nonumber\\
\rho _{gm}&=r_{gm}\exp [{\rm i}(\omega _S-\omega _{3}-\omega _{2}-
\omega _{gm})t]+R_{gm}\exp [{\rm i}(\omega _{1}-\omega _{gm})t]\label{stas},\\
\rho _{gf}&=r_{gf}\exp [{\rm i}(\omega _S-\omega -\omega _{gf})t]+R_{gf}\exp
[{\rm i}(\omega _{1}+\omega _{2}+\omega _{3}-\omega -\omega _{gf})t],
\nonumber\\
\rho _{gk}&=r_{gk}\exp [{\rm i}(\omega _S-\omega _{3}-\omega -\omega
_{gk})t]+R_{gk}\exp [{\rm i}(\omega _{1}+\omega _{2}-\omega -\omega
_{gk})t]. \nonumber
\end{align}
\noindent By substituting  (\ref{stas}) into (\ref{mp}), one can
see that the set of differential equations under consideration
reduces to two independent sets of algebraic equations, where each
refers to the processes determined by only one weak field:
\begin{align}
&{\rm i}R_{gm}D_{gm}=-G_{gm}-R_{gn}G_{nm},\qquad D_{gm}=\Gamma
_{gm}+{\rm i}(\omega _{1}-\omega _{gm}),
\nonumber\\
&{\rm i}R_{gn}D_{gn}=-\int R_{g\varepsilon }G_{\varepsilon
n}d\varepsilon -R_{gm}G_{mn}-R_{gk}G_{kn},\/D_{gn}=\Gamma
_{gn}+{\rm i}(\omega _{1}+\omega _{2}-\omega _{gn}),
\nonumber\\
&{\rm i}R_{g\varepsilon }D_{g\varepsilon }=-R_{gn}G_{n\varepsilon
}-R_{gf}G_{f\varepsilon },\qquad D_{g\varepsilon }={\rm i}(\omega
_{1}+\omega _{2}+\omega _{3}-\omega _{g\varepsilon }),
\label{sis1}\\
&{\rm i}R_{gf}D_{gf}=-\int R_{g\varepsilon }G_{\varepsilon
f}d\varepsilon -R_{gk}G_{kf},\qquad D_{gf}=\Gamma _{gf}+{\rm
i}(\omega _{1}+\omega _{2}+\omega _{3}-\omega -\omega _{gf}),
\nonumber\\
&{\rm i}R_{gk}D_{gk}=-(R_{gn}G_{nk}-R_{gf}G_{fk}),\qquad
D_{gk}=\Gamma _{gk}+{\rm i}(\omega _{1}+\omega _{2}-\omega -\omega
_{gk}), \nonumber
\end{align}
\begin{align}
&{\rm i}r_{g\varepsilon }p_{g\varepsilon }=-G_{g\varepsilon
}-r_{gn}G_{n\varepsilon }-r_{gf}G_{f\varepsilon },\qquad
p_{g\varepsilon }={\rm i}(\omega _S-\omega _{g\varepsilon }),
\nonumber\\
&{\rm i}r_{gn}p_{gn}=-\int r_{g\varepsilon }G_{\varepsilon
n}d\varepsilon -r_{gm}G_{mn}-r_{gk}G_{kn},\/p_{gn}=\Gamma
_{gn}+{\rm i}(\omega _S-\omega _{3}-\omega _{gn}),
\nonumber\\
&{\rm i}r_{gm}p_{gm}=-r_{gn}G_{nm},\qquad p_{gm}=\Gamma _{gm}+{\rm
i}
(\omega _S-\omega _{3}-\omega _{2}-\omega _{gm}),\label{sis2}\\
&{\rm i}r_{gf}p_{gf}=-\int r_{g\varepsilon }G_{\varepsilon
f}d\varepsilon -r_{gk}G_{kf},\qquad p_{gf}=\Gamma _{gf}+{\rm
i}(\omega _S-\omega -\omega _{gf}),
\nonumber\\
&{\rm i}r_{gk}p_{gk}=-r_{gn}G_{nk}-r_{gf}G_{fk},\qquad
p_{gk}=\Gamma _{gk}+{\rm i}(\omega _S-\omega _{3}-\omega -\omega
_{gk}). \nonumber
\end{align}
\noindent Here and later the repeated index $k$ indicates
summation over all discrete off-resonant levels combined to form
the level $k$.
\par
Equations (\ref{sis1}) describe the absorption of ${ E}_{1}$ and
generation at the frequency $\omega _S$, whereas (\ref{sis2})
presents the absorption of ${ E}_{S}$ and parametric conversion of
$E_S$  back into $E_1$. One can solve (12) by substituting the
third equation into the second and fourth. Then, in the
calculation of the integrals,  one can employ  the $\zeta
$-function,
\begin{align}
[-{\rm i}(\omega _S-\omega _{\varepsilon g})]^{-1}=\pi \delta
(\omega _S-\omega _{\varepsilon g})+{\rm i}{\cal P}(\omega
_S-\omega _{\varepsilon g})^{-1}\label{15},
\end{align}
\noindent where $\delta (\xi)$ is the delta function, and $\cal P$
is the principal value obtained by integration. This leads to the
following equations:
\begin{align}
%\begin{align}
&R_{g\varepsilon }={\rm i}[G_{n\varepsilon }-G_{f\varepsilon }
(\gamma _{nf}/\gamma _{ff})\beta _{f}(1-{\rm i}q_{nf})/X_f]
R_{gn}/D_{g\varepsilon },\nonumber\\%\label{16},
%\end{align}
%\begin{align}
&R_{gm}={\rm i}\frac{G_{gm}+ R_{gn}G_{nm}}{X_{m}\Gamma_{gm}},\>%\label{17},
%\end{align}
%\begin{align}
R_{gn}=-\frac{G_{gm}G_{mn}X_f}{ \Gamma_{gf}\Gamma_{gn}X_{m}(1+g_{nn})(X_n
X_f -K+A_{m}X_f)}, \label{18}
%\end{align}
\end{align}
\noindent where
\begin{align}
\beta _{f}&=g_{ff}/(1+g_{ff}),\quad \beta _{n}=g_{nn}/(1+g_{nn}),\quad
 g_{nn}=\gamma_{nn}/\Gamma_{gn},
\quad g_{ff}=\gamma_{ff}/\Gamma_{gf},
\nonumber\\
K&=\beta_f\beta_n(1-{\rm i}q_{nf})^2,\quad A_m=g_{mn}/X_m(1+g_{nn}),\quad
g_{mn}=|G_{mn}|^2/\Gamma_{gm}\Gamma_{gn};\nonumber\\
q_{ij}&=\delta _{ij}/\gamma _{ij},\> \gamma _{ij}=\pi \hbar G_{i\varepsilon
}G_{\varepsilon j}\mid_{\varepsilon =\hbar \omega
_S}+{\rm Re}(G_{ik}G_{kj}/p_{gk}),\nonumber\\
\delta _{ij}&=\hbar P\int d\varepsilon \frac{G_{i\varepsilon
}G_{\varepsilon j}}{(\hbar \omega _S-\varepsilon)} +{\rm
Im}[G_{ik}G_{kj}/p_{gk}],\nonumber\\
%\end{align}
%\begin{align}
X_i&=1+{\rm i}x_i, \>
x_n=(\Omega_1+\Omega_2-\delta_{nn})/(\Gamma_{gn}+\gamma_{nn}),\>
x_m=\Omega_1/\Gamma_{gm},\nonumber\\
x_f&=(\Omega_1+\Omega_2-\Omega_L-\delta_{ff})/(\Gamma_{gf}+\gamma_{ff})
=(\omega_1+\omega_2+\omega_3-\omega-\omega_{gf}-\delta_{ff})/(\Gamma_{gf}+\gamma_{ff}),
\nonumber\\
\Omega_L&=(\omega+\omega_{gf}) -(\omega_3+\omega_{gn}),\>
\Omega_1=\omega_1-\omega_{gm} ,\quad
\Omega_2=\omega_2-\omega_{mn}\label{21}.
\end{align}
%\noindent
Here $\Omega_L$ is the spacing between the quasi-levels induced by
the radiations $E$ and $E_3$ in the continuum.  The Fano
parameters $q_{ij}$ \cite{fano} are assumed real and indicate the
ratio of the light-induced shifts and the broadening of the
corresponding resonances by the control fields. In the adopted
approximation, these parameters are independent of the field
intensities and are governed solely by the properties of the
investigated atom.

Following the same procedure as above and bearing in mind the
condition $\omega_S=\omega_1+\omega_2+\omega_3$, one finds from
the set of equations (\ref{sis2}) that
\begin{align}
r_{g\varepsilon }&={\rm i}\{G_{g\varepsilon }-G_{f\varepsilon }(\gamma
_{gf}/\gamma _{ff})\beta _{f}(1-{\rm i}q_{gf})/X_{f}+ r_{gn}[G_{n\varepsilon
}-G_{f\varepsilon }(\gamma _{nf}/\gamma _{ff})\beta _{f}(1-{\rm
i}q_{nf})/X_{f}]\}/p_{g\varepsilon },
\nonumber\\%\label{23},
r_{gm}&={\rm i}\frac{r_{gn}G_{nm}}{X_{m}\Gamma_{gm}},\>%\label{24},
r_{gn}=\frac{(1-{\rm i}q_{fn})(1-{\rm
i}q_{gf})\gamma_{gf}\gamma_{fn}/(\Gamma_{gf}+\gamma_{ff}) -(1-{\rm
i}q_{gn})X_f\gamma_{gn}}{(1+g_{nn})\Gamma_{gn}(X_f X_n-K+A_m
X_f)}\label{25}.
\end{align}
The calculation of $R_{gm}$ from (\ref{18}) and $r_{g\varepsilon}$
from (\ref{25}), and application of formulas (\ref{pnl}) and
(\ref{10}) after integration over the continuum states, gives the
following expressions for the absorption  and refractive indices
at the frequencies $\omega_{1}$ and $\omega_S$, respectively:
\begin{align}
\dfrac{\alpha (\omega_1)}{\alpha_{10}}&={\rm Re} F_1, \>
\dfrac{n(\omega_1)-1}{2(n_{1 \max}-1)}={\rm Im} F_1,\nonumber\\
F_1&= \dfrac{1}{X_m}{\biggl [}1-\dfrac{A_{m} X_f}
{X_n X_f +A_{m} X_f-K}{\biggr ]},&\label{31n}\\
%\nonumber\\
\dfrac{\alpha(\omega_S)}{\alpha_{S0}}&={\rm Re} F_S,\>
\dfrac{n(\omega_S)-1}{2(n_{S \max}-1)}={\rm Im} F_S,\nonumber\\
F_S&=\biggl [1-\dfrac{X_f X_n(A_n+A_f)-U+A_m A_f X_f}{{X_f X_n-K+A_m X_f}}
\biggr ]
\nonumber\\
&=\biggl [1-A_f-A_n-\dfrac{K(A_n+A_f)-U-A_m A_n X_f}{X_f X_n-K+A_m
X_f}\biggr ]\nonumber\\
&=\biggl [1-A_f-\widetilde A_n-\dfrac{K(\widetilde A_n+A_f)-U}
{X_f \widetilde X_n-K}\biggr ].%&
\label{32}
\end{align}
Here $\alpha _{10}$ is the absorption index with the fields
intensities and detunings turned to zero, and  $\alpha _{S0}$ is
the similar quantity for a transition to the continuum; $n_{1
\max}$ and $n_{S \max}$ are the maximum values of the
corresponding refractive indices under control fields turned off;
and
$$A_f=\beta_f{(1-{\rm i}q_{gf})^2/X_f};\>  A_n=\beta_n{(1-{\rm i}q_{gn})^2/X_n};\>
U=2\beta_f\beta_n(1-{\rm i}q_{gf})(1-{\rm i}q_{fn})(1-{\rm i}q_{ng}).$$%,
The functions $\widetilde A_n=\beta_{n}(1-{\rm
i}q_{gn})^2/\widetilde X_n$ and $\widetilde X_n=X_n+A_m$ account
for the perturbation of a two-photon resonance with the level $n$
by the strong fields. The expressions for the refractive index are
obtained on the assumption that this index is close to unity in
the absence of the fields.
\par
The calculation of $R_{g\varepsilon}$ and $r_{gm}$ from the
expressions (\ref{18}) and (\ref{25}), which is carried out making
use of formulas (\ref{pnl}) and (\ref{10}) and integration over
the states in the continuum, gives  expressions for the FWM
nonlinear susceptibility  at $\omega _S =\omega _1 +\omega
_2+\omega _3$ as
\begin{align}
\frac{\chi^{(3)}(\omega_S)}{\chi_{S0}^{(3)}}=
\frac{X_f-\beta_f(1-{\rm i}q_{nf})(1-{\rm i}q_{fg})/(1-{\rm
i}q_{ng})}{X_m(1+g_{nn}) (X_n X_f-K+A_m X_f)}, \label{27}
\end{align}
\noindent and for the nonlinear susceptibility determining
 conversion of the $E_S$ radiation back into the
radiation of frequency $\omega _{1}$ as
\begin{align}
\frac{\chi^{(3)}(\omega_1)}{\chi_{10}^{(3)}}= \frac{X_f-\beta_f(1-{\rm
i}q_{nf})(1-{\rm i}q_{fg})/(1-{\rm i}q_{ng})}{X_m(1+g_{nn}) (X_n X_f-K+A_m
X_f)} \label{30}
\end{align}
\noindent In the expressions (\ref{27}) and (\ref{30}), the
quantities $\chi ^{(3)}_{S0}$ and $\chi ^{(3)}_{10}$ are
fully-resonant nonlinear susceptibilities for non-perturbative
weak fields.
%%%%%%%%%%%%%%%5
\subsection{Laser-induced structures in absorption and refraction\\
spectra at discrete and continuous transitions}

In this subsection, we
analyze and  compare the effects of strong fields on discrete and
continuous spectra with the aid of the derived formulas.  An
important difference is seen as compared with  similar effects if
solely discrete transitions are involved  \cite{Sok,Vved}. The
results given below demonstrate considerable perturbations of
discrete spectra by the radiations coupled to the continuum. The
first term in (\ref{31n}) represents the field-unperturbed
absorption coefficient for the $gn$ transition, and the second
term refers to the cumulative effects of the strong fields.  The
coefficient $A_m$ represents splitting of a resonance into two
components by the strong field $E_2$ \cite{Sok,Vved} and modified
by the strong field $E_3$. The function  $K$ describes
modification of these effects by the strong field $E$. It is
proportional to the product of the intensities of the fields $E$
and $E_3$. The effects in question disappear when any of these
fields is turned off. Since the field $E_2$ is in resonance with a
discrete transition and the fields $E_3$ and $E$ are coupled to
the continuum, the spectral properties of the corresponding
contributions are different. If $E_3 = 0$ ($\gamma_n=0, \beta_n
=0, K =0, g_{nn}=0$), the equation for absorption in (\ref{31n})
converges to the standard one for three-level nonlinear
spectroscopy \cite{Sok,Vved},
\begin{equation}
{\alpha(\omega_1)/\alpha_{01}}=\Re\left[X_n\left/\left(X_n X_m
+\frac{g_{mn}}{1+g_{nn}}\right)\right.\right]. \label{35}
\end{equation}

\begin{figure}[!h]
\begin{center}
\includegraphics[width=0.5\textwidth]{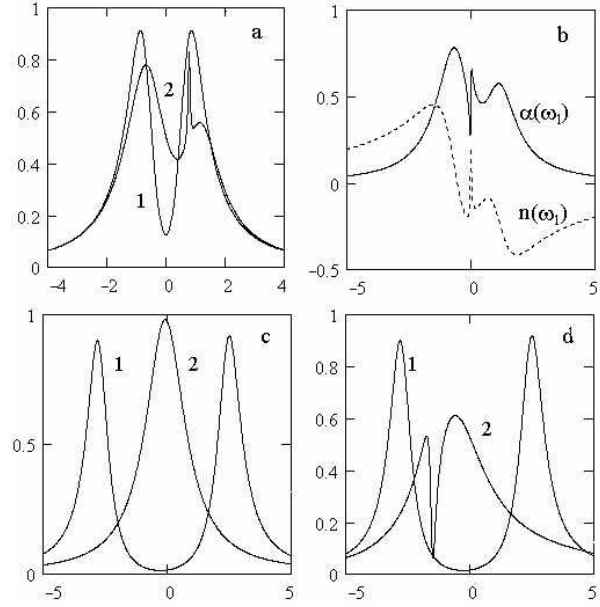}\end{center}
\caption[recon] {\label{abs1} Absorption index at $\omega_1$
(reduced by its resonant value in the absence of all strong
fields) vs one-photon detunings $\Omega_1/\Gamma_{gm}$. Here,
$q_{ff}=0.9$, $q_{nn}=0.5$, $\Gamma_{gm}/\Gamma_{gf}=100$, and
$\Gamma_{gm}/\Gamma_{gn}=10$. {\bf (a,b)}: $\Omega_2=0$,
$q_{fn}=1.5$, $g_{mn}=7$, $\gamma_{ff}/\Gamma_{gf}=2$. {\bf (a)}:
$\gamma_{nn}=0 (1)$, $\gamma_{nn}/\Gamma_{gn}=5$,
$\Omega_L/\Gamma_{gm}=0.8 (2)$. {\bf (b)}:
$\gamma_{nn}/\Gamma_{gn}=5$, $\Omega_L=0$. {\bf (c,d)}:
$\Omega_2/\Gamma_{mn}=0.3$, $g_{mn}=70$,
$\gamma_{ff}/\Gamma_{gf}=10$, $\Omega_L/\Gamma_{gm}=-1.1$,
$\gamma_{nn}=0 (1)$, $\gamma_{nn}/\Gamma_{gn}=50 (2)$. {\bf (c)}:
$q_{fn}=15$. {\bf (d)}: $q_{fn}=1.5$.}
\end{figure}
The factor in the parentheses in the above formula has two
singularities with respect to $x_m$, which indicates splitting of
the resonance into two maxima. The positions of these maxima and
their relative amplitudes may vary depending on the parameters of
the fields and transitions. Resonance splitting is stipulated by
the appearance of coherence at the  transition $ng$ and by the
consequent appearance of additional quantum transitions in which
photons of frequency $\hbar\omega_1$ may participate. The phase
and relaxation properties of the corresponding term in nonlinear
polarization at the frequency $\omega_1$ are represented by the
dispersion function $X_n$. The additional strong fields $E_3$ and
$E$ perturb the quantum system, which, as pointed out earlier,
causes additional modification of the laser-induced quasi-levels
and resonances. The changes in the refractive index can he
explained similarly.

The spectral characteristics of absorption at the frequency
$\omega_S$ are governed by the interference between three quantum
pathways: directly to the continuum, to level $f$, and also to the
superposition of levels $n$ and $m$. The influence of the first
two processes was investigated earlier in Refs.
\cite{GP2,GP3,Kn,Vved} and is accounted for by the function
$\widetilde A_l$ in the equations (\ref{32}). The strong fields
$E_3$ and $E_2$ lead to additional changes in the spectra. The
additional independent structure, described in these expressions
by the function $\widetilde A_n$, is supplemented by the
interference structures proportional to the functions $K$ and $U$,
which disappear when any of the fields $E_3$ or $E$ is turned off
or when the spacing between the quasi-levels $\Omega_L$ is
increased.

Figure {\ref{abs1}} shows the dependence of the absorption index
at $\omega_1$ on the scaled detunings from the bare-state
one-photon resonance $\Omega_1/\Gamma_{gm}$. The dashed plot at
Fig. {\ref{abs1}}(b) shows  the lineshape of the dispersive index
at the frequency $\omega_1$  for the same parameters. The plot
indicates very strong dispersion, induced by the dressing fields
coupled to both discrete and continuum states. Figures
{\ref{abs1}}(c) and {\ref{abs1}}(d) illustrate the sensitivity of
the spectra, induced jointly by two dressing fields $E_3$ and $E$,
on the Fano parameter $q_{fn}$. One can create transparency in
certain frequency intervals of the discrete transitions or, on the
contrary, eliminate effects of the dressing fields with the aid of
destructive interference by varying the intensities and detunings
of the driving fields as well as the coupled levels.

\begin{figure}[!h]
\begin{center}
\includegraphics[width=0.35\textwidth]{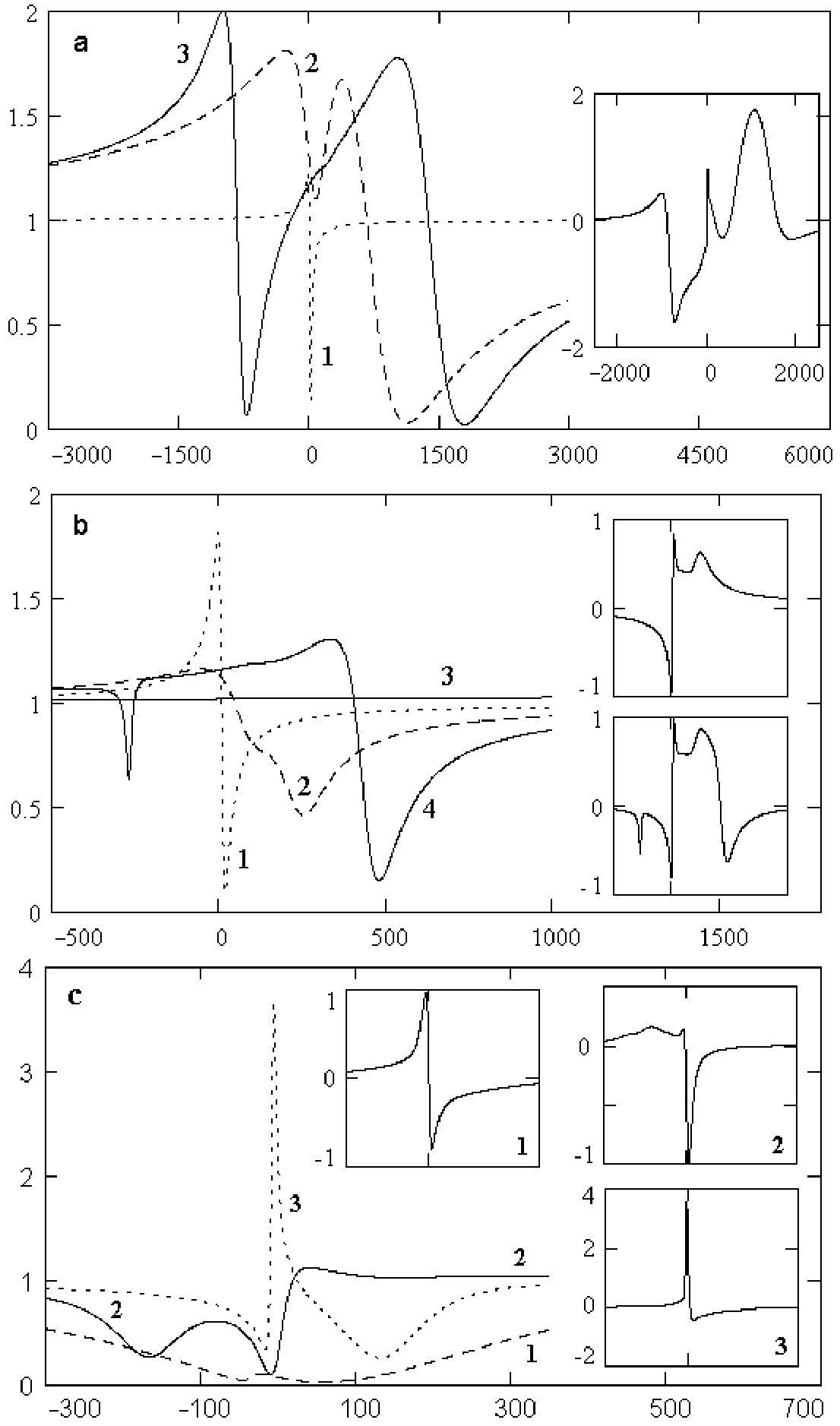}\hspace{5mm}
\includegraphics[width=0.35\textwidth]{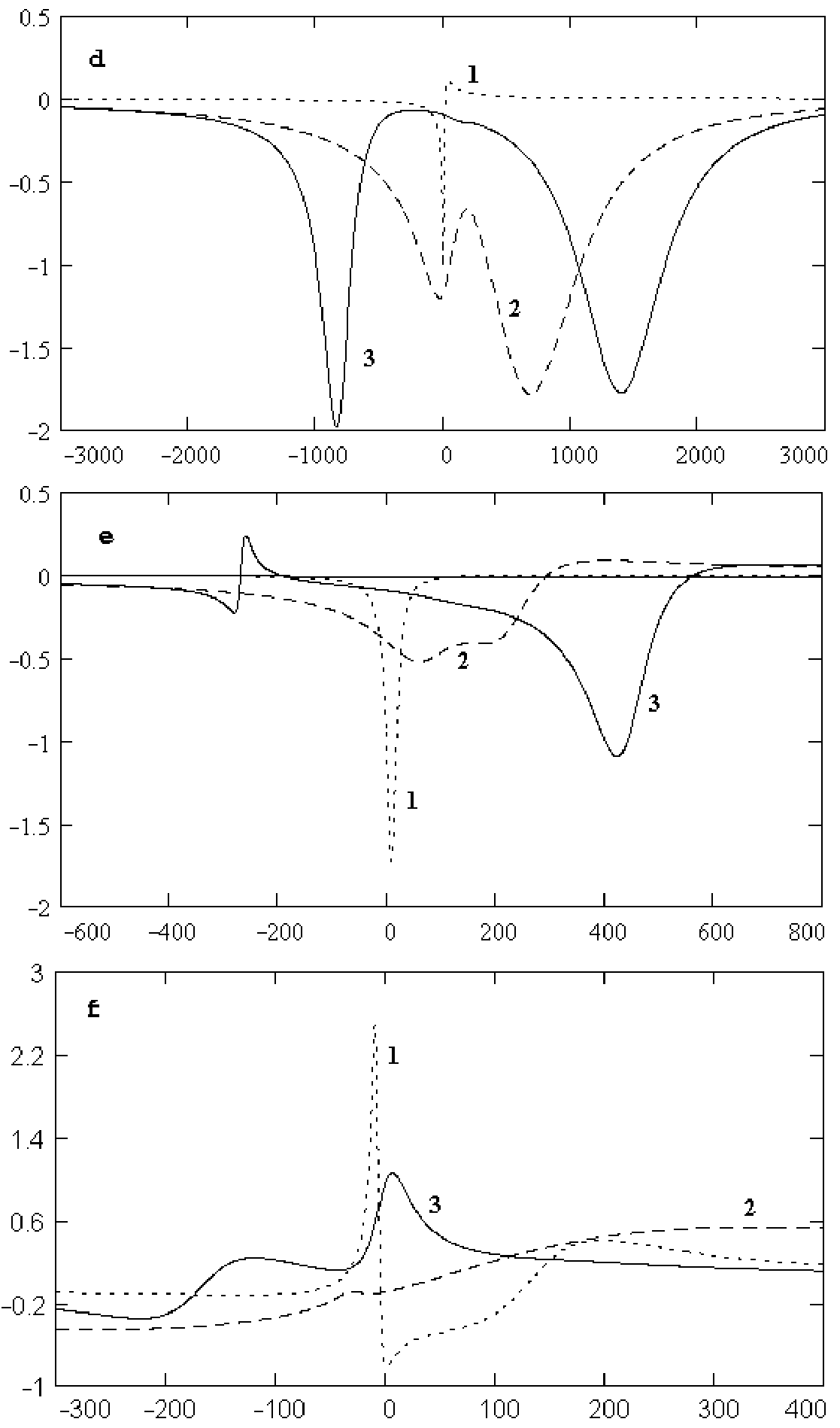}
\end{center}
\caption[recon] {\label{abs4} Absorption index at $\omega_{S}$
(reduced by its value in the absence of all strong fields) vs
detunings
$\Omega/\Gamma_{fg}=(\omega_S-\omega-\omega_{fg})/\Gamma_{fg}$
{\bf (a-c)}. The inserts are the interference contributions to the
corresponding curves vs detuning (for {\bf (b,c)} the detuning
interval is the same as for the main curves). Here
$\Gamma_{gm}/\Gamma_{gf}=100$, $\Gamma_{gm}/\Gamma_{gn}=10$. {\bf
(a,b)}: $q_{ff}=0.9$, $q_{nn}=0.9$, $\Omega_L/\Gamma_{gf}=-110$,
and $\Omega_2/\Gamma_{gf}=30$. {\bf (a)}: $q_{gf}=-0.5$,
$q_{gn}=-0.95$, $q_{fn}=15$. {\bf (b)}: $q_{gf}=-0.95$,
$q_{gn}=-0.5$, $q_{fn}=15$(1,2,3), $q_{fn}=150$(4). {(1)}
$E_2=E_3=0$, $\gamma_{ff}/\Gamma_{gf}=10$. {(2-4)} $g_{mn}=70$,
$\gamma_{nn}/\Gamma_{gn}=50$. {(2)} $\gamma_{ff}=0$. {(3,4)}
$\gamma_{ff}/\Gamma_{gf}=10$. {\bf (c)}: $q_{gf}=0.95$,
$q_{gn}=0.01$, $q_{ff}=0.01$, $q_{nn}=-5$, $q_{fn}=1.5$,
$\Omega_2/\Gamma_{gf}=0$, $\gamma_{ff}/\Gamma_{gf}=10$,
$\Gamma_{gf}=-1530$. {(2,3)} $\gamma_{nn}/\Gamma_{gn}=5$. {(2)}
$\Omega_L/\Gamma_{gf}=-110$. {(3)} $\Omega_L/\Gamma_{gf}=-405$.
{\bf (d-f)}: Scaled dispersive profile of LICS at $\omega_S$ vs
$\Omega/\Gamma_{fg}$. (All parameters are the same as in plots
{\bf (a-c)}, consequently.) In the absence of the driving fields
absorption level corresponds to 1, dispersive -- to zero.}
\end{figure}

Figures {\ref{abs4}}(a) and {\ref{abs4}}(b) show the dependence of
the absorption index at the frequency $\omega_S$ (reduced by its
value in the absence of the dressing fields) on the scaled
detuning
$\Omega/\Gamma_{fg}=(\omega_S-\omega-\omega_{fg})/\Gamma_{gf}$.
The plots indicate the sensitivity of the probe field absorption
on the Fano parameters  (in the considered cases on $q_{gf}$ and
$q_{gn}$) and demonstrate new possibilities for  manipulation of
LICS with extra dressing fields coupling adjacent transitions.
Interference contributions, which disappear in the absence of
either the $E$ or $E_3$ field, are shown in the inputs to the
figure. Figure {\ref{abs4}}(c) demonstrates the  sensitivity of
the absorption spectrum on the strength of the driving field $E_3$
as well as on the spacing of two quasi-levels, induced in the
continuum (on two-photon detuning $\Omega_L=\omega-\omega_3-
\omega_{fn}$). The figures show possible manipulation of {LICS} in
the absorption index for a probe radiation (including formation of
transparency windows). This is  brought about by the interference
of two {LICS} (quantum pathways via discreet and continuum
states), induced by the fields at $\omega_3$ and $\omega$ and
modified by the strong field at $\omega_2$.

The refractive index is proportional to the derivative of the
structures over frequency. The plots {\ref{abs4}(d,e,f)} display
the potential of manipulating the magnitude and lineshape of the
factor $(n(\omega_S)-1)/(n_{S\max}-1)$ with the change of
intensities and detunings of the dressing fields, as well as
strong dependence on the Fano parameters, which may find
applications in short-wavelength optics.

The nonlinear interference effects involving quantum transitions
may influence differently the linear and nonlinear
susceptibilities. Therefore, under certain conditions a reduction
in the absorption of the initial and generated radiations can be
simultaneously followed by an increase in the nonlinear
polarization caused by the resonant effects and by the
constructive interference in the strong fields. It is thus
possible to increase considerably the power of the generated
short-wavelength radiation.

\subsection{Resonance sum-frequency generation in strongly-absorbing
media enhanced by quantum interference}\label{gl}

In weak fields the nonlinear susceptibility increases strongly
upon approaching to discrete resonances, but this is accompanied
by enhancement of the absorption of the initial radiations.
Depending on the detunings from the resonances, on the ratio of
the oscillator strengths of the transitions, and on the radiation
intensities, either the absorption of the initial radiation $E_1$
or of the generated radiation may predominate. A numerical
analysis of the influence of these factors on changes in the
frequency dependence of the power of the generated
short-wavelength radiation during the course of propagation in an
optically dense medium can be made if the quantum efficiency of
conversion given by expression (\ref{qe}) is rewritten in the
form:
\begin{equation}
\eta_{\rm q}(z)=\frac{4\overline\eta_{{\rm q}0}}{|\overline b|} \exp\big
[-(\overline\alpha_1+C\overline\alpha_S){z_0}\big ] \bigg\{{\rm sinh}^2\bigg
[\sqrt{{(|\overline b|-\overline b)C/ 2}}\cdot{z_0}\bigg ]+\sin ^{2}
\bigg[\sqrt{{(\mid \overline b\mid+\overline
b)C/2}}\cdot{z_0}\bigg]\bigg\},\label{qen}
\end{equation}
where $\overline \eta_{{\rm q}0} = \tilde\eta_{{\rm q}0}
/(\alpha_{10}\alpha_{S0}), \> \overline b=b
/(\alpha_{10}\alpha_{S0})= 4\overline\eta_{{\rm
q}0}-(\overline\alpha_{1}-C\overline\alpha_{S})^2/4C, \>
\overline\alpha_1=\alpha_1/\alpha_{10}; \>
\overline\alpha_S=\alpha_S/\alpha_{S0},\>
 C=\alpha_{S0}/\alpha_{10}, \> and z_0=z\alpha_{10}/2$.

The expression for $\overline\eta_{{\rm q}0}$, describing the
quantum efficiency of conversion over a distance
$1/\sqrt{\alpha_{10}\alpha_{S0}}$, considered within the
approximation of  ignoring absorption of given fields, becomes
\begin{equation}
\overline\eta_{{\rm q}0}= \eta_{{\rm q}0}^0 |\overline \chi^{(3)}|^2
g_{mn}g_{nn},
\end{equation}
where $\eta_{{\rm q}0}^0$ is the quantum efficiency of conversion
based on the resonant unperturbed nonlinearity over a distance
$1/\sqrt{\alpha_{10}\alpha_{S0}}$ in fields corresponding to
$g_{mn}=g_{nn}=1; \> \overline\chi^{(3)}=\chi^{(3)}/\chi_0$. We
shall henceforth use the following approximate expressions:
\begin{eqnarray}
\alpha_{10}=4\pi {\omega_1}{|d_{gm}|^2/ c \hbar\Gamma_{gm}},\>
\alpha_{S0}=4\pi^2{(\omega_S/ c)}|d_{g\varepsilon}|^2\bigg
|_{\varepsilon=\hbar \omega_3},
\nonumber\\
|\chi_0|^2=(\pi/2\hbar^2)^2(1+q_{gn}^2)|d_{gm}d_{mn}d_{n\varepsilon}d_{\varepsilon
g}|^2 (\Gamma_{gm}\Gamma_{gn})^{-2}.
\end{eqnarray}
In this approximation the factor $\eta_{{\rm q}0}^0$ is determined
completely by the Fano parameter $q_{gn}$:
\begin{equation}
\eta_{{\rm q}0}^0=k_1' k_S' |2\pi \chi_0|^2|\big /
d_{mn}d_{n\varepsilon}|^2\pi(16\hbar^3
\Gamma_{gm}\Gamma_{gn}^2)^{-1}\alpha_{10}\alpha_{S0}=(1+q_{gn}^2).
\label{etaf}
\end{equation}

As pointed our earlier, the laser-induced absorption and
transparency resonances of the radiation $E_1$ can be interpreted
as splitting of the level $m$ into quasi-levels by the strong
field $E_2$ (in combination with the fields $E_3$ and $E$). The
laser-induced resonances of the generated radiation are determined
by two quasi-levels in the continuum and appear near the
frequencies $\omega_{ng}+\omega_3$ and $\omega_{fg}+\omega$. These
quasi-levels are separated by an energy $\hbar\Omega_L$. The
detuning of the generated radiation frequency from the first
resonance amounts to $\Omega_S=\Omega_1+\Omega_2$,  and that from
the second resonance is $\Omega=\Omega_S-\Omega_L$. The
corresponding quantum channels may interfere differently when the
detunings $\Omega_1,\>\Omega_2$ and $\Omega_L$ are altered, which
is manifested in the spectral characteristics of the absorption
and nonlinear-optical generation processes. The relative role of
these channels is governed by the intensities of the radiations,
by their detunings from the resonances, and by the ratio of the
oscillator strengths for the $g-m$ and $g-\varepsilon$
transitions. We shall now illustrate these relationships by
considering several numerical models.

\begin{figure}[!h]
\begin{center}
\includegraphics[width=0.6\textwidth]{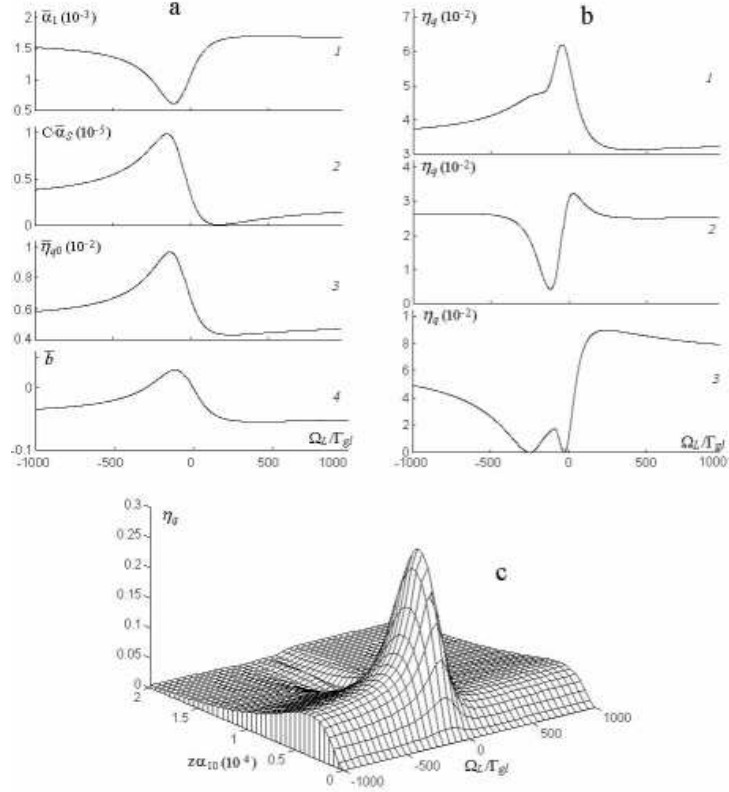}
%\hspace{15mm}
\end{center}
\caption{Dependence of the absorption indices $\alpha_1 /
\alpha_{10}$ and  $\alpha_S/\alpha_{10}$, of the value
$\overline\eta_{{\rm q}0}$ (that is proportional to the squared
modulus of the amplitude of the nonlinear polarization at the
entrance to the medium), and of the conversion rate $\overline b$
on the detuning $\Omega_L$ {\bf(a)}. Dependence of the quantum
conversion efficiency $\eta_{\rm q}$ on $\Omega_L$ for
$z\alpha_{10}=8.5\cdot 10^3$ (1), $z\alpha_{10}=1\cdot 10^4$ (2),
and $z\alpha_{10}=2\cdot 10^4$ (3) {\bf (b)}. Dependence of the
quantum conversion efficiency $\eta_{\rm q}$ on the optical
thickness and on $\Omega_L$ {\bf (c)}. Here, $C=10^{-5}$,
$g_{ff}=150$, $g_{nn}=200$, $g_{mn}=9000$,
$\Omega_1/\Gamma_{gf}=5000$, $\Omega_2/\Gamma_{gf}=-5100$,
$q_{fg}=0.95$, $q_{gn}=-2$, $q_{ff}=0.01$, $q_{nn}=-5$,
$q_{fn}=0$, $\Gamma_{gm}/\Gamma_{gf}=100$,
$\Gamma_{gm}/\Gamma_{gn}=10$ {\bf(a-c)}.} \label{pic2}
\end{figure}%\vspace{-5mm}

Figure \ref{pic2} illustrates the case when the integral
oscillator strength for a transition to an energy interval $E$ (of
the order of the level widths) in the continuum is considerably
less than for the $gm$ transition $(C= 10^{-5})$. The frequencies
$\omega_1$ and $\omega_2$ are detuned far from their one-photon
transitions, but the frequency sum is close to a perturbed
two-photon resonance. It is evident from Fig. \ref{pic2} that
changes in $\Omega_L$ (because of variation of the frequencies
$\omega$ or $\omega_3$) reduce, for the selected Fano parameters
and radiations, the absorption coefficient $\alpha_1$ by
approximately threefold, increase $\alpha_S$ approximately by a
factor of 3.8 in one detuning interval, and reduce considerably
this coefficient in the other interval, whereas the square of the
modulus of the nonlinear polarization (proportional to
$\overline\eta_{{\rm q}0}$) increases by a factor of 1.9 (these
changes are relative to the values of the corresponding parameters
in the far wings where the effects of the strong field $E$ do not
appear) [Fig. \ref{pic2} (a)]. The absorption coefficient for a
transition to the continuum remains on the whole much less than
$\alpha_1$ ($\alpha_S/\alpha_1 =
C\overline\alpha_S/\overline\alpha_1 \approx 10^{-2}$). In a
certain range of $\Omega_L$, the sign of $\overline b $ becomes
positive, and the nonlinear-optical conversion rate begins to
exceed the rate of absorption of the radiation. It is in this
interval that there is a sharp maximum of the quantum efficiency
of conversion (the dependence of this conversion efficiency on the
optical thickness and on  $\Omega_L$ is demonstrated in Fig.
\ref{pic2} (c), where the maximum $\eta_{{\rm q}\max}=0.29$ is
reached for $z \alpha_{10}=4000$). The absorption of the radiation
$E_1$ alters considerably the dependence of the power of the
generated radiation on $\Omega_L$ along the medium [Fig.
\ref{pic2} (b)].

\begin{figure}[!h]
\begin{center}
\includegraphics[width=0.6\textwidth]{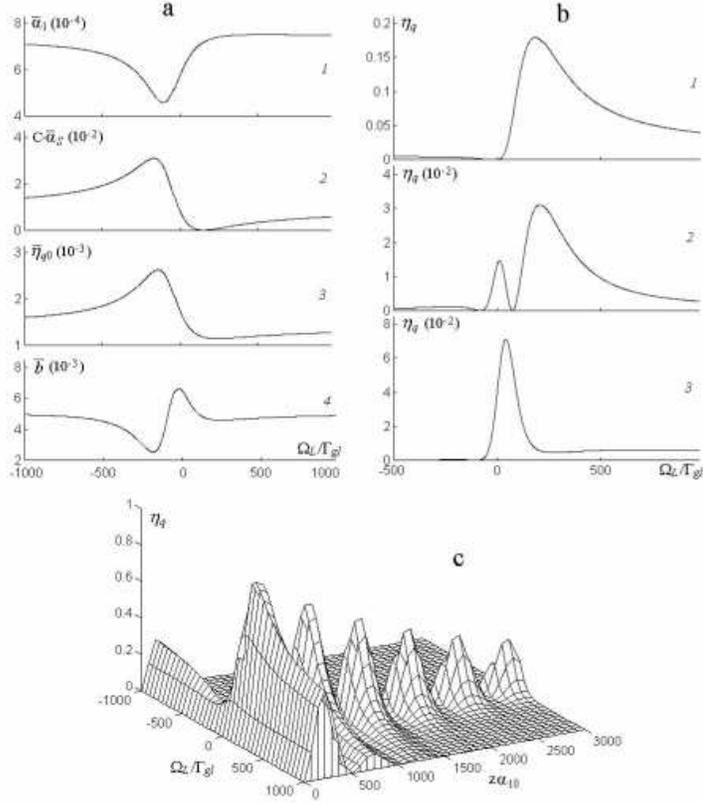}
\end{center}
\caption{Absorption indices $\alpha_1/\alpha_{10}$ and
$\alpha_S/\alpha_{10}$, the value $\overline\eta_{{\rm q}0}$ and
conversion rate $\overline b$ plotted as a function of $\Omega_L$
{\bf (a)}. Dependence of the quantum conversion efficiency
$\eta_{\rm q}$ on $\Omega_L$ along the medium computed for
$z\alpha_{10}=4.5\cdot 10^2$ {(1)}, $z\alpha_{10}=5\cdot 10^2$
(2), and $z\alpha_{10}=5.5\cdot 10^2$ {(3)} {\bf (b)}. Dependence
of $\eta_{\rm q}$ on the optical thickness and on $\Omega_L$ {\bf
(c)}. Here, $C=3\cdot 10^{-2}$, $g_{nn}=500$, $g_{mn}=8000$ {\bf
(a-c)}. The other parameters are the same as in Fig. \ref{pic2}.}
\label{pic3}
\end{figure}

Figure \ref{pic3} is computed for the case where the detuning from
a one-photon resonance is still large, but the oscillator
strengths for a discrete transition and a transition to the
continuum differ less $(C=3\cdot 10^{-2})$. Then, under the action
of the field $E$, variation of $\Omega_L$ can reduce the
absorption coefficient $\alpha_1$ by a factor of just 1.5, whereas
the absorption coefficient $\alpha_S$ increases approximately
threefold in one interval, but falls considerably in the other
interval; practically throughout the whole range of the detuning
$\Omega_L$, the dominant effect is the absorption transition into
the continuum ($\alpha_S/\alpha_1 \approx 70$), and
$\overline\eta_{{\rm q}0}$ increases by a factor of 1.9 [Fig.
\ref{pic3}(a)]. The quantity $\overline b$ is always positive and
has a characteristic maximum in a certain interval of $\Omega_L$.
For a given detuning, the rate of conversion begins to exceed the
rate of absorption so much that an oscillatory regime appears
along the medium. Propagation in an optically-dense medium makes
the spectrum of the output power significantly dependent on the
optical length of the medium, i.e., on $z$ or on the concentration
$N$ of atoms. The absolute amplitude and the position of the
maximum change, and new resonances appear [Fig. \ref{pic3}(b)]. At
the first maximum (corresponding to $z\alpha_{10}\approx 125$) the
quantum efficiency of conversion can reach 0.9 [Fig.
\ref{pic3}(c)], which is three times greater than in the preceding
case.

\begin{figure}[!h]
\begin{center}
\includegraphics[width=0.5\textwidth]{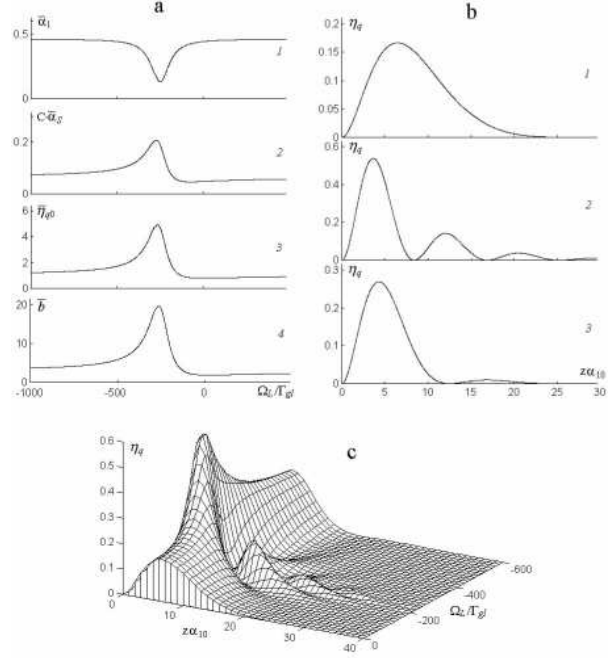}
\end{center}
\caption{Dependence of the absorption indices
$\alpha_1/\alpha_{10}$ and $\alpha_S/\alpha_{10}$, of the value
$\overline\eta_{{\rm q}0}$, and of the conversion rate $\overline
b$ on $\Omega_L$ {\bf (a)}. Dependence of the quantum conversion
efficiency $\eta_{\rm q}$ on the optical thickness for
$\Omega_L/\Gamma_{gf}=0$ {(1)}, $\Omega_L/\Gamma_{gf}=-250$ {(2)},
$\Omega_L/\Gamma_{gf}=-400$ {(3)} {\bf (b)}. Dependence of the
quantum conversion efficiency $\eta_{\rm q}$ on the optical
thickness and on $\Omega_L$ {\bf (c)}. Here, $g_{ff}=100$,
$g_{nn}=5$, $g_{mn}=7$, $\Omega_1/\Gamma_{gf}=0$,
$\Omega_2/\Gamma_{gf}=-250$ {\bf (a-c)}. The other parameters are
the same as in Fig. \ref{pic3}.} \label{pic4}
\end{figure}

In view of the nonresonant nature of the interaction, the
attainment of such high values of the quantum efficiency of
conversion requires high intensities of the $E_2$ and $E_3$,
radiations, and also long lengths (or high densities of atoms) of
the medium. The use of resonant processes makes it possible to
reduce the required intensities and to reach considerable values
of the quantum efficiency by optimization of the bleaching and
interference effects. Figure \ref{pic4} illustrates the case where
the ratio of  oscillator strengths is the same as in Fig.
\ref{pic3}, but there is no detuning from a one-photon resonance.
At intensities of the $E_2$ radiation three orders of magnitude
less than in the preceding case, it is possible to reduce the
absorption coefficient of the $E_1$ radiation approximately by a
factor of 10 compared with the value of this coefficient in the
absence of strong fields (the maximum effect of the field $E$ is a
reduction in this coefficient by a factor of 1.5) [Fig.
\ref{pic4}(a)]. In the region of reduction in $\alpha_1$, the
value of $\alpha_S$ increases approximately threefold and
$\overline\eta_{{\rm q}0}$ by a factor of 4.7. The absorption
coefficients $\alpha_1$ and $\alpha_S$ are then comparable, and
the nonlinear-optical conversion rate exceeds considerably, in a
wide range of $\Omega_L$, the rates of absorption of the
radiations, reaching a sharp maximum at
$\Omega_L/\Gamma_{gf}=-250$. This gives rise to an oscillatory
regime of generation along the medium in this detuning range
[Figs. \ref{pic4}(b,c)], i.e., it leads to the feasibility of
total (apart from that lost by absorption) conversion of the
radiations $E_1$ and $E_S$ (and vice versa) for certain products
of the length of the medium and the concentration of atoms. For
the selected parameters, the conversion efficiency at the first
maximum (where $z\alpha_{10}\approx 5$) reaches 0.54. This is less
than in the preceding case because of the much greater absolute
values of the absorption coefficient, but this efficiency is
reached at much lower intensities and for much smaller lengths of
the nonlinear medium and much lower concentration of atoms in this
medium.

\subsection{Absorption and dispersive spectra
at Doppler-broadened transitions}
\begin{figure}[!h]\begin{center}
\includegraphics[width=0.7\textwidth]{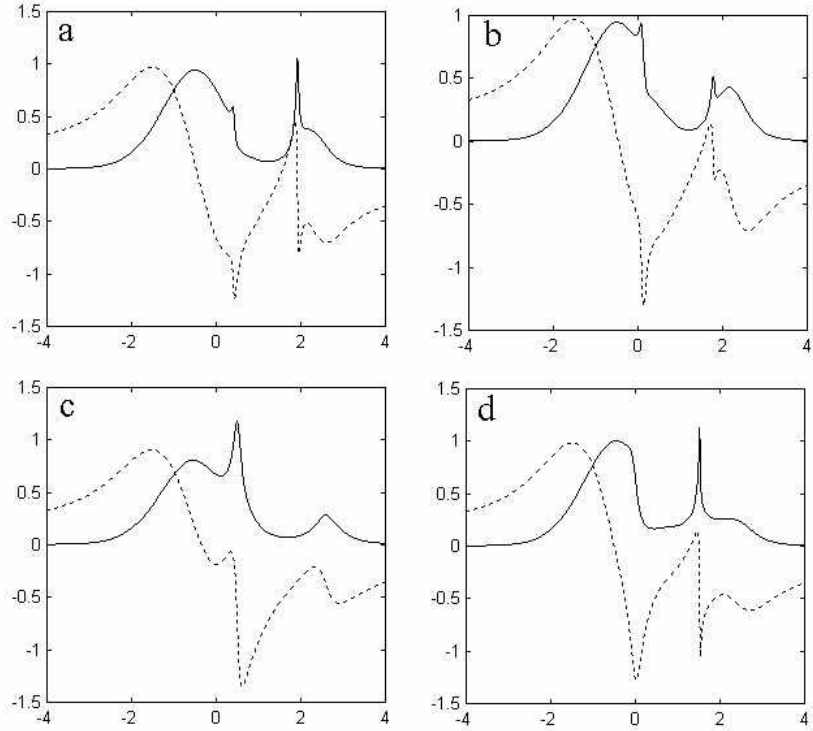}\end{center}
\caption[recon] {\label{fiad}  Absorption (solid) and refractive
(dash) indices at $\omega_1$ (reduced by its value in the absence
of all strong fields) in  a Doppler-broadened medium vs detunings
$\Omega_1/\Delta\omega_{1D}$. Here, the Doppler HWHM
$\Delta\omega_{1D}=16.65 \Gamma_{gm}$, the wavevector orientations
are ${\bf k}\uparrow\uparrow {\bf k}_1$, ${\bf
k}_2\uparrow\uparrow {\bf k}_3 \uparrow\downarrow {\bf k}_1$, and
$k_2/k_1=0.9$, $k_3/k_1=0.5$, $k/k_1=0.6$ (${\bf k}_i$ is
wave-vector corresponding to the frequency $\omega_i$),
$\Gamma_{gm}/\Gamma_{gf}=100$, $\Gamma_{gm}/\Gamma_{gn}=10$;
$|G_{mn}|^2/(\Delta\omega_{1D})^2=1$, $q_{nn}=0.5$, $q_{ff}=0.9$,
$\Omega_2/\Delta\omega_{1D}=9$. {\bf (a,b)}:
$\gamma_{nn}/\Delta\omega_{1D}=0.2$,
$\gamma_{ff}/\Delta\omega_{1D}=0.1$, $q_{fn}=0.5$. {\bf (a)}:
$\Omega_L=0$.
%{\bf b)}: $\Omega_L/\Delta\omega_{1D}=0$.
{\bf (b-d)}: $\Omega_L/\Delta\omega_{1D}=-0.8$. {\bf (c)}:
$\gamma_{nn}/\Delta\omega_{1D}=0.8$,
$\gamma_{ff}/\Delta\omega_{1D}=0.3$, $q_{fn}=-1.5$. {\bf (d)}:
$\gamma_{nn}/\Delta\omega_{1D}=0.2$,
$\gamma_{ff}/\Delta\omega_{1D}=0.8$, $q_{fn}=1.5$.}
\end{figure}
%%%%%%%%%%%%%%%%%%%%%%%%%%%%%%%%%
%%%%%%%%%%%%%%%%%%%%%%%%%%%%%%%%%%%%%
\begin{figure}[!h]\begin{center}
\includegraphics[width=0.7\textwidth]{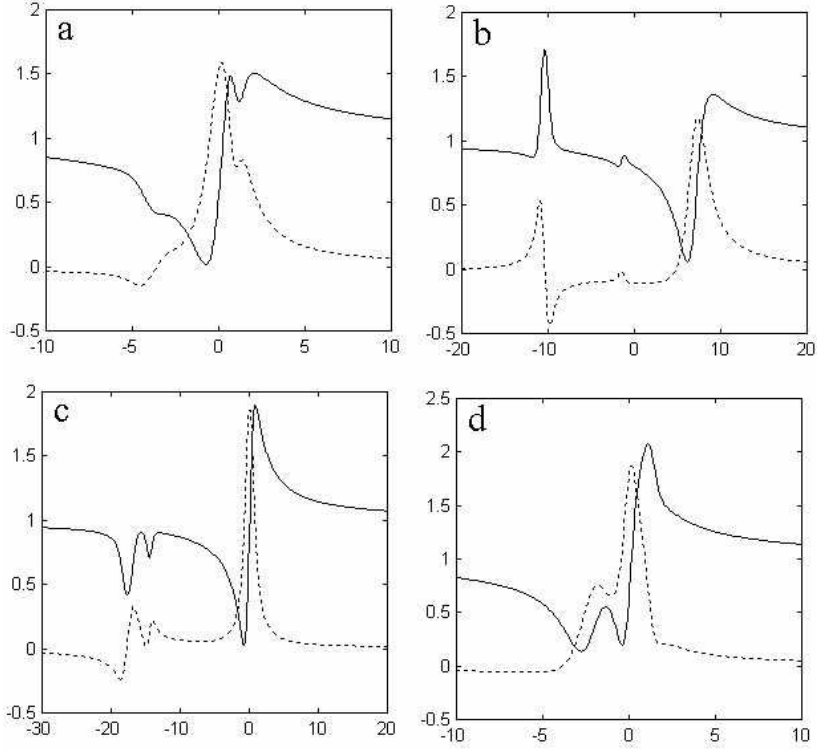}\end{center}
 \caption[recon] {\label{fibd} Absorption (solid) and
refractive (dash) indices at $\omega_S$ (reduced by its value in
the absence of all strong fields) at Doppler-broadened transitions
vs detunings
$\Omega/\Delta\omega_{4D}=(\omega_S-\omega-\omega_{fg})/\Delta\omega_{SD}$.
Here, the Doppler HWHM $\Delta\omega_{SD}=5\cdot 10^3
\Gamma_{gf}$; the wavevector orientations are ${\bf
k}\uparrow\uparrow {\bf k}_3\uparrow\uparrow {\bf k}_2
\uparrow\uparrow {\bf k}_4$, and $k/k_4=0.8$, $k_3/k_4=0.3$,
$k_2/k_4=0.37$, $\Gamma_{gm}/\Gamma_{gf}=100$,
$\Gamma_{gm}/\Gamma_{gn}=10$;
$|G_{mn}|^2/(\Delta\omega_{SD})^2=1$,
$\gamma_{nn}/\Delta\omega_{SD}=0.4$,
$\gamma_{ff}/\Delta\omega_{SD}=0.8$, $q_{gf}=0.95$, $q_{gn}=0.01$,
$q_{ff}=0.01$, $q_{nn}=-5$. {\bf (a)}: $q_{fn}=1.5$,
$\Omega_L/\Delta\omega_{SD}=1.5$,
$\Omega_2/\Delta\omega_{SD}=2.2$. {\bf (b-d)}: $\Omega_2=0$. {\bf
(b)}: $q_{fn}=15$, $\Omega_L/\Delta\omega_{SD}=1.5$. {\bf (c)}:
$q_{fn}=-1.5$, $\Omega_L/\Delta\omega_{SD}=15$. {\bf (d)}:
$q_{fn}=-1.5$, $\Omega_L/\Delta\omega_{SD}=-0.5$.}
\end{figure}

A Lorentzian line profile with near natural linewidth can be
observed only by making use special techniques, like atomic jets.
Analysis shows that the contributions of atoms at different
velocities in a Doppler-broadened medium may completely change the
features  described above. Figures \ref{fiad} and \ref{fibd}
demonstrate that proper adjustment of  the orientation of the
wavevectors, along with the intensities and detunings of the
coupled waves, provides additional means to  manipulate the
lineshape. Moreover, enhanced subDoppler structures can be formed
by the compensation for Doppler shifts by power shifts. For
details of the physics, see \cite{Sh,Ba} and references therein.
The plots dshow that the interference of contributions from the
atoms at different velocities brings an important distinction in
appearance of quantum interference processes at coupled discrete
and continuous states. The figures show that the appearance of
quantum interference at Doppler-broadened transitions also may be
both constructive and distractive, depending on the detunings of
$\omega-\omega_3$ from $\omega _{nf}$, on the Fano parameters, on
the detunings from the two-photon resonance $gn$, on the ratios of
the wavenumbers, and on orientations of the wavevectors.
%%%%%%%%%%%%%%%%%%%
%%%%%%%%%%%%%%%%%%%%
\section{Four-wave mixing, dissociation and population transfer
controlled by  LICS in folded energy-level schemes}

Here, we will
consider the folded scheme, which is characteristic for  molecules
with low-lying predissociation states. Such a coupling
configuration allows coherent laser control of dissociation
through the 'dark' states that are not connected with the ground
one by the allowed transition. On the other hand, the scheme under
consideration enables one to transfer the population between two
upper bound states which are not connected directly too. Thus, we
will explore the possibilities of manipulating dissociation and
populations of excited states through implementation of the
interference of quantum pathways via a variety of low-lying
continuum states.

\begin{figure}[!h]
\begin{center}\vspace{1cm}
\includegraphics[width=0.5\textwidth]{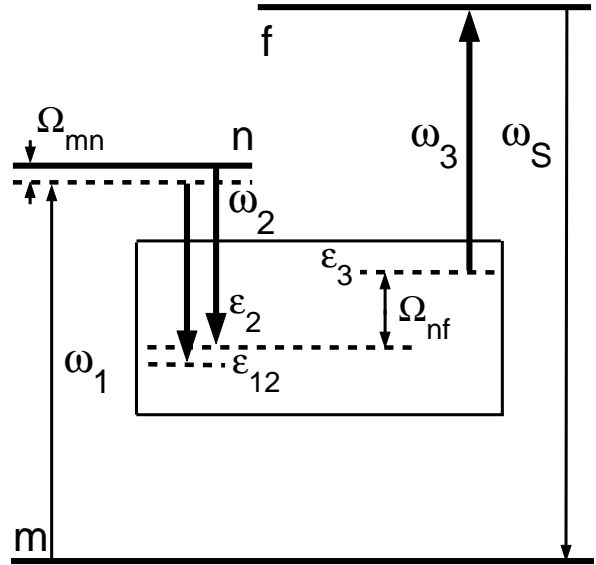}
\end{center}
\caption{\label{levb} LICS-based coherent control in the folded
schemes.}
\end{figure}

The proposed coupling scheme is illustrated in the energy level
diagram depicted in Fig. \ref{levb}. The radiation at frequency
$\omega_1$ couples the bound-bound transition $m-n$, and
radiations at the frequencies $\omega_2$ and $\omega_3$ couple the
bound states $n$ and $f$ with the states of dissociation continuum
$\varepsilon$, as shown in the picture. Density matrix equations,
which are employed  for investigating population transfer, account
for both relaxation and incoherent excitation of the discrete
states:
\begin{align}
&\dot\rho_{n\varepsilon}=i\left(\rho_{nn}V_{n\varepsilon}+
\rho_{nf}V_{f\varepsilon}\right),\nonumber\\
&\dot \rho_{f\varepsilon}=i\left(\rho_{ff}V_{f\varepsilon}+
\rho_{fn}V_{n\varepsilon}\right),\nonumber\\
&\dot \rho_{m\varepsilon}=i\left(\rho_{mn}V_{n\varepsilon}+
\rho_{mf}V_{f\varepsilon}\right),\nonumber\\
&\dot \rho_{mn}+\Gamma_{mn}\rho_{mn}=i\left(\rho_{mm}V_{mn}+\int
d\varepsilon\rho_{m\varepsilon}
V_{\varepsilon n}-V_{mn}\rho_{nn}\right),\nonumber\\
&\dot \rho_{mf}+\Gamma_{mf}\rho_{mf}=i\left(\int
d\varepsilon\rho_{m\varepsilon}
V_{\varepsilon f}-V_{mn}\rho_{nf}\right),\label{off}\\
&\dot \rho_{nf}+\Gamma_{nf}\rho_{nf}=i\left(\int
d\varepsilon\rho_{n\varepsilon} V_{\varepsilon f}-\int d\varepsilon V_{n
\varepsilon}\rho_{\varepsilon f}
-V_{nm}\rho_{mf}\right),\nonumber\\
&\dot\rho_{\varepsilon\varepsilon}=-2\Re\left(i\int d\varepsilon
V_{\varepsilon n}\rho_{n\varepsilon}+i\int d\varepsilon V_{\varepsilon
f}\rho_{f\varepsilon}\right),\nonumber
\end{align}
\begin{align}
&\dot \rho_{nn}+\Gamma_{n}\rho_{nn}=-2\Re\left(i\int d\varepsilon
V_{n\varepsilon}
\rho_{\varepsilon n}+iV_{nm}\rho_{mn}\right)+Q_n,\nonumber\\
&\dot \rho_{ff}+\Gamma_{f}\rho_{ff}=-2\Re\left(i\int d\varepsilon
V_{f\varepsilon}
\rho_{\varepsilon f}\right)+Q_f,\label{po}\\
&\dot\rho_{mm}+\Gamma_{m}\rho_{mm}=-2\Re\left(iV_{mn}\rho_{nm}\right)+Q_m+
w_{nm}\rho_{nn}. \nonumber
\end{align}
Here, $Q_{i}$ is the rate of incoherent excitation to level $i$,
$\Gamma_{i}\rho_{ii}$ is the rate of relaxation, $w_{nm}$ is the
probability of excitation of level $m$ due to the relaxation
transitions from level $n$, and $V_{jk}$ and $V_{j\varepsilon}$
are the corresponding matrix elements of the quasi-resonant
electrodipole interaction of the driving field and molecules
scaled to Planck's constant $\hbar$. The system of equations
(\ref{po}) corresponds to the case of an {\it open} energy level
configuration. This implies that all the levels coupled with the
driving fields, including level $m$, can be incoherently excited
from the reservoir of much greater populated lower levels, so that
the parameters $Q_j$ are constant.

For the {\it closed} schemes, i.e., the schemes where level $m$ is
a ground one, and therefore incoherent excitation of the upper
levels depends on population transfer stimulated by the applied
fields, the equations for the populations take the form
\begin{align}
&\dot \rho_{ff}+\Gamma_{f}\rho_{ff}=-2\Re\left(i\int d\varepsilon
V_{f\varepsilon}
\rho_{\varepsilon f}\right)+w_f\rho_{mm},\nonumber\\
&\dot \rho_{nn}+\Gamma_{n}\rho_{nn}=-2\Re\left(i\int d\varepsilon
V_{n\varepsilon}
\rho_{\varepsilon n}+iV_{nm}\rho_{mn}\right)+w_n\rho_{mm},\label{pc}\\
&\rho_{mm}=1-\rho_{nn}-\rho_{ff}-\int\dot W dt. \nonumber
\end{align}
Here, $w_i$ describes the probability of excitation to level $i$
from the ground state, and $\dot W$ is rate of dissociation.

In the equations (\ref{off}), the terms like $\rho_{\varepsilon
k}V_{kj}$ ($j,k=n,m$) are discarded. This is valid because
coherence at discrete-discrete two-photon transitions is usually
much stronger than that at bound-free two-photon transitions.  The
driving fields couple the interval of continuum states in the
vicinity of the resonant energy $\varepsilon_0$ that is roughly
equal to the width of the coupled power-broadened discrete levels.
However, the oscillation strength at a  bound-free transition is
distributed over a much wider energy interval. Therefore, the
fraction of oscillator strength attributed to the coupled energy
interval is usually relatively small. The requirement derived with
the aid of the Laplace transformation indicates that the indicated
terms in the above equations can be neglected until
\begin{align}
&{\partial f}/{\partial \varepsilon_{\varepsilon=\varepsilon_0}}>>\hbar
|V_{jk}|{\partial^2 f}/{\partial \varepsilon^2_{\varepsilon=
\varepsilon_0}},&\label{f}
\end{align}
where $f(\varepsilon)$ is the energy density of the transition
oscillator strength. Such a requirement is fulfilled practically
for all realistic atomic and molecular continua in the energy
range that is well above the ionization or dissociation threshold.

\subsection{Quasistationary solution of density matrix  equations}\label{qsc}

In this section we shall assume that all radiations are continuous
waves or rectangular pulses with duration $\tau$ that is much
greater than all relaxation times in the quantum system under
investigation. We will consider also the case of small loss of
molecules due to dissociation for the period of measurement or the
pulse duration, so that the quasi-stationary regime is
established. The corresponding requirement is
\begin{align}
\int\limits_0^\tau\dot Wdt=W(\tau)<<1,\label{usl1}
\end{align}
where the dissociation yield $\dot Wdt$ is described by the equation
\begin{align}
\dot W=\dot \rho_\varepsilon=-2\Re\left(i\int d\varepsilon V_{\varepsilon n}
\rho_{n\varepsilon}+i\int d\varepsilon V_{\varepsilon
f}\rho_{f\varepsilon}\right).\label{diss}
\end{align}
Quasi-stationary solutions of the density matrix can be found in
the form
\begin{align}
&\rho_{ii}=r_i\quad (i=m,n,f),\quad
\rho_{ij}=r_{ij}\exp(i\Omega_{ij}t),\quad
\rho_{j\varepsilon}=r_{j\varepsilon}\exp(i\Omega_{j\varepsilon}t),\nonumber\\
&V_{ij}=G_{ij}\exp(i\Omega_{ij}t),\quad
V_{j\varepsilon}=G_{j\varepsilon}\exp(i\Omega_{j\varepsilon}t).\label{stat}
\end{align}
Here,  $G_{ij}=E_kd_{ij}/2\hbar$,
$G_{j\varepsilon}=E_kd_{j\varepsilon}/2\hbar$,
$\Omega_{mn}=\omega_1-\omega_{mn}$,
$\Omega_{nf}=\omega_3-\omega_2-\omega_{nf}$,
$\Omega_{mf}=\omega_1-\omega_2+\omega_3-\omega_{mf}$,
$\Omega_{n\varepsilon}=\omega_{2}-\omega_{n\varepsilon}$,
$\Omega_{f\varepsilon}=\omega_{3}-\omega_{f\varepsilon}$, and
$\Omega_{m\varepsilon}=\omega_{1}-\omega_{2}-\omega_{m\varepsilon}$.
The amplitudes of the resonant fields $E_k$ (see Fig. \ref{levb}),
and consequently all amplitudes $r_{i}$, $r_{ij}$, and
$r_{j\varepsilon}$, are assumed to be independent of time. Then,
by introducing the expressions (\ref{stat}) to the equations
(\ref{off}), we can reduce them to a system of algebraic ones:
\begin{align}
&ir_{n\varepsilon}p_{n\varepsilon}=-r_nG_{n\varepsilon}-r_{nf}G_{f\varepsilon}
%+G_{nm}r_{m\varepsilon}
,\quad%\nonumber\\
p_{n\varepsilon}=i(\omega_2-\omega_{n\varepsilon});\nonumber\\
&ir_{f\varepsilon}p_{f\varepsilon}=-r_{f}G_{f\varepsilon}-r_{fn}G_{n\varepsilon},
\quad%\nonumber\\
p_{f\varepsilon}=i(\omega_3-\omega_{f\varepsilon});
\nonumber\\
&ir_{m\varepsilon}p_{m\varepsilon}=-r_{mn}G_{n\varepsilon}-r_{mf}G_{f\varepsilon}
%+G_{mn}r_{n\varepsilon}
,\quad%\nonumber\\
p_{m\varepsilon}=i(\omega_1-\omega_2-\omega_{m\varepsilon});\nonumber\\
&ir_{mn}p_{mn}=-\int d\varepsilon r_{m\varepsilon}G_{\varepsilon n}-r_m
G_{mn}+G_{mn}r_n,
\quad%\nonumber\\
p_{mn}=\Gamma_{mn}+i(\omega_1-\omega_{mn});%\nonumber
\label{statsys}\\
&ir_{mf}p_{mf}=-\int d\varepsilon r_{m\varepsilon}G_{\varepsilon
f}+G_{mn}r_{nf},
\quad%\nonumber\\
p_{mf}=\Gamma_{mf}+i(\omega_1-\omega_2+\omega_3-\omega_{mf});\nonumber\\
&ir_{nf}p_{nf}=-\int d\varepsilon r_{n\varepsilon}G_{\varepsilon f}+\int
d\varepsilon G_{n\varepsilon}r_{\varepsilon f}+G_{nm}r_{mf},\quad
%\nonumber\\
p_{nf}=\Gamma_{fn}\pm i(\omega_3-\omega_2-\omega_{nf}).\nonumber
\end{align}
Further by introducing the solutions of the first, second and
third equations to the integrals and with the aid of the $\zeta
$-function (\ref{15}), we obtain
\begin{align}
&r_{mn}\left[\delta_{nn}^m-\Omega_{mn}+i(\Gamma_{mn}+\gamma_{nn}^m)\right]=
\left(r_n-r_m\right)G_{mn}-r_{mf}\left(\delta_{fn}^m+i\gamma_{fn}^m\right),\nonumber\\
&r_{mf}\left[\delta_{ff}^m-\Omega_{mf}+i(\Gamma_{mf}+\gamma_{ff}^m)\right]=G_{mn}r_{nf}-
r_{mn}\left(\delta_{nf}^m+i\gamma_{nf}^m\right),\label{sys3}\\%\nonumber\\
&r_{nf}\left[\delta_{ff}^n-\delta_{nn}^f-
\Omega_{nf}+i(\Gamma_{nf}+\gamma_{ff}^n+\gamma_{nn}^f)\right]=%\nonumber\\
G_{nm}r_{mf}+r_f\left(\delta_{nf}^f-i\gamma_{nf}^f\right)-
r_{n}\left(\delta_{nf}^n+i\gamma_{nf}^n\right),\nonumber
\end{align}
where
\begin{align}
&\gamma_{nn}^m=\pi\hbar G_{n\varepsilon_{12} }G_{\varepsilon_{12} n}
+\Re\left(G_{nk}G_{kn}/p_{mk}\right),\nonumber\\
&\delta_{nn}^m=\hbar{\cal P}\int d\varepsilon \cdot G_{n\varepsilon
}G_{\varepsilon
n}/(\varepsilon_{12}-\varepsilon)+\Im\left(G_{nk}G_{kn}/p_{mk}\right);\nonumber\\%%%%%%%%%%%%
&\gamma_{ff}^m=\pi \hbar G_{f\varepsilon_{12} }G_{\varepsilon_{12} f}
+\Re\left(G_{fk}G_{kf}/p_{mk}\right),\nonumber\\
&\delta_{ff}^m=\hbar{\cal P}\int d\varepsilon \cdot G_{f\varepsilon
}G_{\varepsilon
f}/(\varepsilon_{12}-\varepsilon)+\Im\left(G_{fk}G_{kf}/p_{mk}\right);\nonumber\\%%%%%%%%%%%%
&\gamma_{nf}^m=\pi \hbar G_{n\varepsilon_{12} }G_{\varepsilon_{12} f}
+\Re\left(G_{nk}G_{kf}/p_{mk}\right),\\%\nonumber\\
&\delta_{nf}^m=\hbar{\cal P}\int d\varepsilon \cdot G_{n\varepsilon
}G_{\varepsilon
f}/(\varepsilon_{12}-\varepsilon)+\Im\left(G_{nk}G_{kf}/p_{mk}\right);\nonumber\\%%%%%%%%%%%%%%
&\gamma_{fn}^m=\pi \hbar G_{f\varepsilon_{12} }G_{\varepsilon_{12} n}
+\Re\left(G_{fk}G_{kn}/p_{mk}\right),\nonumber\\
&\delta_{fn}^m=\hbar{\cal P}\int d\varepsilon \cdot G_{f\varepsilon
}G_{\varepsilon
n}/(\varepsilon_{12}-\varepsilon)+\Im\left(G_{fk}G_{kn}/p_{mk}\right);\nonumber \label{gdel}%\\%%%%%%%%%%%%%%%
\end{align}
\begin{align}
&\gamma_{nn}^n=\pi \hbar G_{n\varepsilon_{2}}G_{\varepsilon_{2} n}
+\Re\left(G_{nk}G_{kn}/p_{kn}\right),\nonumber\\
&\delta_{nn}^n=\hbar{\cal P}\int d\varepsilon \cdot G_{n\varepsilon
}G_{\varepsilon
n}/(\varepsilon_{2}-\varepsilon)+\Im\left(G_{nk}G_{kn}/p_{kn}\right);\nonumber\\%%%%%%%%%%%%%%%%%%
&\gamma_{ff}^n=\pi \hbar G_{f\varepsilon_2 }G_{\varepsilon_2 f}
+\Re\left(G_{fk}G_{kf}/p_{kn}\right),\nonumber\\
&\delta_{ff}^n=\hbar{\cal P}\int d\varepsilon \cdot G_{f\varepsilon
}G_{\varepsilon
f}/(\varepsilon_2-\varepsilon)+\Im\left(G_{fk}G_{kf}/p_{kn}\right);\nonumber\\%%%%%%%%%%%%%%%%
&\gamma_{nf}^n=\pi \hbar G_{n\varepsilon_2 }G_{\varepsilon_2 f}
+\Re\left(G_{nk}G_{kf}/p_{kn}\right),\\%\nonumber\\
&\delta_{nf}^n=\hbar{\cal P}\int d\varepsilon \cdot G_{n\varepsilon
}G_{\varepsilon
f}/(\varepsilon_2-\varepsilon)+\Im\left(G_{nk}G_{kf}/p_{kn}\right);\nonumber\\%%%%%%%%%%%%%%%%%%
&\gamma_{fn}^n=\pi \hbar G_{f\varepsilon_2 }G_{\varepsilon_2 n}
+\Re\left(G_{fk}G_{kn}/p_{kn}\right),\nonumber\\
&\delta_{fn}^n=\hbar{\cal P}\int d\varepsilon \cdot G_{f\varepsilon
}G_{\varepsilon
n}/(\varepsilon_2-\varepsilon)+\Im\left(G_{fk}G_{kn}/p_{kn}\right);\nonumber \label{gdel1}%\\%%%%%%%%%
\end{align}
\begin{align}
&\gamma_{nn}^f=\pi \hbar G_{n\varepsilon_3 }G_{\varepsilon_3 n}
+\Re\left(G_{nk}G_{kn}/p_{kf}\right),\nonumber\\
&\delta_{nn}^f=\hbar{\cal P}\int d\varepsilon \cdot G_{n\varepsilon
}G_{\varepsilon
n}/(\varepsilon_3-\varepsilon)+\Im\left(G_{nk}G_{kn}/p_{kf}\right);\nonumber\\%%%%%%%%%%%
&\gamma_{ff}^f=\pi \hbar G_{f\varepsilon_3 }G_{\varepsilon_3 f}
+\Re\left(G_{fk}G_{kf}/p_{kf}\right),\nonumber\\
&\delta_{ff}^f=\hbar{\cal P}\int d\varepsilon \cdot G_{f\varepsilon
}G_{\varepsilon
f}/(\varepsilon_3-\varepsilon)+\Im\left(G_{fk}G_{kf}/p_{kf}\right);\nonumber\\%%%%%%%%%%%%%%%%%%%%
&\gamma_{nf}^f=\pi \hbar G_{n\varepsilon_3 }G_{\varepsilon_3 f}
+\Re\left(G_{nk}G_{kf}/p_{kf}\right),\\%\nonumber\\
&\delta_{nf}^f=\hbar{\cal P}\int d\varepsilon \cdot G_{n\varepsilon
}G_{\varepsilon
f}/(\varepsilon_3-\varepsilon)+\Im\left(G_{nk}G_{kf}/p_{kf}\right);\nonumber\\%%%%%%%%%%%%%%%%%
&\gamma_{fn}^f=\pi \hbar G_{f\varepsilon_3 }G_{\varepsilon_3 n}
+\Re\left(G_{fk}G_{kn}/p_{kf}\right),\nonumber\\
&\delta_{fn}^f=\hbar{\cal P}\int d\varepsilon \cdot G_{f\varepsilon
}G_{\varepsilon
n}/(\varepsilon_3-\varepsilon)+\Im\left(G_{fk}G_{kn}/p_{kf}\right).\nonumber %\label{gdel2}
\end{align}
and
\begin{align}
\varepsilon_{12}&=\hbar(\omega_1-\omega_2),\quad
\varepsilon_{2}=E_n-\hbar\omega_2,\quad
\varepsilon_{3}=E_f-\hbar\omega_3,\\
p_{mk}&=\Gamma_{mk}+i(\omega_1-\omega_2-\omega_{mk}),\quad
p_{kn}=\Gamma_{kn}+i(\omega_{kn}-\omega_2),\quad
p_{kf}=\Gamma_{kf}+i(\omega_{kf}-\omega_3).\nonumber
\end{align}
Besides the continuum states, a contribution of other non-resonant
levels $k$ is taken into account too and a sum over repeating $k$
index is assumed. The contribution from these levels may occur
commensurable to that of the continuum states. As seen from the
equations (\ref{sys3}), the values $\gamma_{ij}$ and $\delta_{ij}$
describe light-induced broadening and shifts of discrete
resonances stipulated by the induced transitions between them
through the continuum. The parameters
$q_{ij}=\delta_{ij}/\gamma_{ij}$ are analogous to the Fano
parameters for autoionizing states. Within the validity of
(\ref{f}), their dependence  on the field intensities can be
neglected. These parameters determine the most important features
of the processes under investigation since they characterize the
relative integrated contribution of all off-resonant quantum
states compared to the resonant ones.

With the aid of (\ref{statsys}), the equation for the dissociation
rate (\ref{diss}) takes the form
\begin{align}
\dot W=\dot
r_\varepsilon=2\left[\gamma_{nn}^nr_n+\gamma_{ff}^fr_f+
2\Re(\gamma_{fn}r_{nf})\right].\label{re}
\end{align}
The first two terms on RHS describe induced transitions to the
continuum from the corresponding discrete levels, whose
populations are determined by a variety of processes. Most
important is the last term describing quantum interference, which
occurs if two LICS overlap. Let us consider specific features of
the processes in both open and closed energy level configurations.
%\newpage
\subsubsection{Open configuration}\label{open}
In this case in the same way as above, one obtains
\begin{align}
&r_m\Gamma_m=w_{nm}r_n+Q_m-2\Re(iG_{mn}r_{nm}),\nonumber\\
&r_f(\Gamma_f+2\gamma_{ff}^f)=Q_f-2\Re\left(r_{nf}[\gamma_{fn}^f+
i\delta_{fn}^f]\right),\label{ri}\\
&r_n(\Gamma_n+2\gamma_{nn}^n)=Q_n-2\Re\left(iG_{nm}r_{mn}+r_{nf}[(\gamma_{fn}^n-
i\delta_{fn}^n]\right).\nonumber
\end{align}
Then the solution of (\ref{sys3}) and (\ref{ri}) can be presented
in the form
\begin{align}
&r_{nf}={\left\{(r_n-r_m)|G_{mn}|^2
\gamma_{nf}^m(1-iq_{nf}^m)-[r_f\gamma_{nf}^f(1+iq_{nf}^f)+
r_n\gamma_{nf}^n(1-iq_{nf}^n)]Y\right\}}Z^{-1},\nonumber\\
&r_m=(L_1C_1+L_2S_1)/(C_1C_2-S_1S_2),\,
r_n= (L_2+ S_2r_m)/C_1,\nonumber\\
&r_f=[Q_f+2r_n B_4^{fn}+2(r_m-r_n)M_4^{fm}]Y_f^{-1}.\label{nasel}
\end{align}
The notations introduced here are as follows:
\begin{align}
&C_1=Y_{nm}Y_f-L_1L_2,\,
C_2=[\Gamma_m+2(G-M_1^{mm})]Y_f-4M_4^{fm}M_3^{mf},\nonumber\\
&S_1=[w_{nm}+2(G+M_1^{mn}-D)]Y_f+2L_2M_3^{mf},\,%\label{csf}
S_2=PY_f+2M_4^{fm}L_1,\nonumber\\
&F_1=Q_m Y_f+2Q_f M_3^{mf},\quad F_2=Q_n Y_f-Q_fL_1,\,
P=2(G-M_1^{nm}+D),\nonumber\\
&L_1=2(B_3^{nf}-2M_3^{mf}), \, L_2=2(B_4^{fn}-M_4^{fm}),\,
Y_f=\widetilde\Gamma_f-2B_2^{ff},\nonumber\\
&Y_{nm}=\widetilde\Gamma_n+2(G-B_1^{nn}+M_1^{mn}-D),\,
G=|G_{mn}|^2\Re(y_{mf}/Y),\nonumber\\
&B_i^{jl}=\gamma_{nn}^j\gamma_{nn}^l\Re(b_i^{jl}Y/Z),\,
M_i^{jl}=|G_{mn}|^2\gamma_{nn}^j\gamma_{ff}^l\Re(b_i^{jl}/Z),\nonumber\\
&D=|G_{mn}|^4\gamma_{nn}^m\gamma_{ff}^m\Re[k_1^{mm}/(YZ)],\,
j,l=m,n,f;\nonumber\\
&Y=y_{mn}y_{mf}-\gamma_{nn}^m\gamma_{ff}^m b_1^{mm},\,
Z=y_{mf}y_{nf}y_{mn}-\gamma_{nn}^m\gamma_{ff}^m
b_1^{mm}y_{nf}+|G_{mn}|^2y_{mn},\nonumber\\
&y_{mn}=\widetilde\Gamma_{mn}+i(\Omega_{mn}-\delta_{nn}^m),\,%\nonumber\\
y_{mf}=\widetilde\Gamma_{mf}+i(\Omega_{mf}-\delta_{ff}^m),\nonumber\\
&y_{nf}=\widetilde\Gamma_{nf}+i(\Omega_{nf}+\delta_{nn}^f-\delta_{ff}^n),\,
b_1^{jl}=k_o(1-iq_{fn}^j)(1-iq_{nf}^l),\nonumber\\
&b_2^{jl}=k_o(1+iq_{fn}^j)(1+iq_{nf}^l),\,
b_3^{jl}=k_o(1-iq_{fn}^j)(1+iq_{nf}^l),\nonumber\\
&b_4^{jl}=k_o(1+iq_{fn}^j)(1-iq_{nf}^l),\,
k_o=(\gamma_{nf}^j\gamma_{fn}^l)/(\gamma_{nn}^j\gamma_{ff}^l),\,
\widetilde\Gamma_{nf}=\Gamma_{nf}+\gamma_{nn}^f+\gamma_{ff}^n,\nonumber\\
&\widetilde\Gamma_{mn}=\Gamma_{mn}+\gamma_{nn}^m,\,
\widetilde\Gamma_{mf}=\Gamma_{mf}+\gamma_{ff}^m,\,
\widetilde\Gamma_{n}=\Gamma_n+2\gamma_{nn}^n+w_{nm},\,
\widetilde\Gamma_{f}=\Gamma_f+2\gamma_{ff}^f.\label{obozn}
\end{align}

First, consider the effect of the interference of two LICS in the
case where the laser field at discrete transition is turned off
($G_{mn}=0$). Then the solution (\ref{nasel}) takes the more
simple form of
\begin{align}
&r_{nf}=-[r_f\gamma_{nf}^f(1+iq_{nf}^f)+ r_n\gamma_{nf}^n(1-iq_{nf}^n)]
y_{nf}^{-1},\,
r_m = Q_m/\Gamma_m+w_{nm}r_n,\nonumber\\
&r_n=(Q_nY_f-2Q_f\gamma_{nn}^n\gamma_{ff}^f\Re\left(b_3^{nf}/y_{nf}\right))/\Delta,\nonumber\\
&r_f=(Q_fY_n+2Q_n\gamma_{nn}^f\gamma_{ff}^n\Re\left(b_4^{fn}/y_{nf}\right))/\Delta,\nonumber\\
&Y_n
=\widetilde\Gamma_n-2\gamma_{nn}^n\gamma_{ff}^n\Re\left(b_1^{nn}/y_{nf}\right),\,
Y_f=\widetilde\Gamma_f-2\gamma_{nn}^f\gamma_{ff}^f\Re\left(b_2^{ff}/y_{nf}\right),\nonumber\\
&\Delta=Y_fY_n-4(\gamma_{nn}^n\gamma_{nn}^f)^2
\Re\left(b_4^{fn}/y_{nf}\right)\Re\left(b_3^{nf}/y_{nf}\right).\label{G0}
\end{align}
The process of dissociation is described by the formula (\ref{re})
with the aid of (\ref{G0}). All the terms proportional to
$\Re(...)$ describe interference processes and disappear for
either of the driving fields $E_2$ or $E_3$ being turned off. The
corresponding spectral structures are power broadened (terms with
$\gamma$) and of an asymmetric shape that is determined by the
power-shifts (terms with $\delta$) and by the Fano parameters $q$
for the corresponding coupled excited levels. The position and
shape of the laser-induced autoionizing-like resonance within the
continuum can be varied with the change of the frequencies and
intensities of the driving fields. Constructive interference can
be turned into destructive by small variation of the overlap of
two LICS, which forms a basis for coherent quantum control of
photophysical processes by the driving lasers. This will be
illustrated below with numerical examples.

If either $E_2$ or $E_3$ is turned off ($\gamma_{nf}^j=0$), the
equations (\ref{nasel}) take the standard form that accounts for
power-broadening of the level $f$ or $n$ depending on which field
is turned off. When $E_3=0$, we obtain
\begin{align}
r_m&=\frac{Q_m+Q_nw_{nm}/\widetilde\Gamma_n+2(Q_m+Q_n)|G_{mn}|^2/|y_{mn}|^2}
{\Gamma_m+2(\Gamma_m+\widetilde\Gamma_n-w_{nm})|G_{mn}|^2/|y_{mn}|^2},\quad
r_f=Q_f/\widetilde\Gamma_f,\nonumber\\
r_n&=\frac{Q_m\Gamma_m/\widetilde\Gamma_n+2(Q_m+Q_n)|G_{mn}|^2/|y_{mn}|^2}
{\Gamma_m+2(\Gamma_m+\widetilde\Gamma_n-w_{nm})|G_{mn}|^2/|y_{mn}|^2},\quad
\dot W=2\gamma_{nn}^n r_n.\label{tpd}
\end{align}
These expressions describe a process of two-photon dissociation
without any interference phenomena in the spectral continuum.
Opportunities for manipulating  dissociation and population
transfer through the interference of two LICS follows from a
comparison of equations (\ref{nasel}) and (\ref{tpd}).
\subsubsection{Closed configuration}\label{closestat}
Assuming  $\int \dot W dt<<1$ in (\ref{pc}), so that
$r_m+r_n+r_f\approx 1$, the equations for levels populations in a
closed scheme can be written as
\begin{align}
r_n\widetilde\Gamma_n&=-2\Re\left[iG_{nm}r_{mn}+r_{nf}
(\gamma_{fn}^n+i\delta_{fn}^n)\right]+w_nr_m,\nonumber\\
r_f\widetilde\Gamma_f&=-2\Re\left[r_{fn}(\gamma_{nf}^f+i\delta_{nf}^f)\right]+w_fr_m,\
r_m=1-r_n-r_f.\label{clr}
\end{align}
Then, with the aid of (\ref{sys3}), the solution for the populations of the
excited levels can be presented in the form
\begin{align}
r_m=&1-r_n-r_f,\quad
r_n=\left[Y_f(w_n+P)+L_1(w_f+2M_4^{fm})\right]{\Delta_c}^{-1},
\nonumber\\
r_f=&\left[Y_{nm}(w_f+2M_4^{fm})+L_2(w_n+P)\right]{\Delta_c}^{-1},\\%\quad
\Delta_c=&(\widetilde\Gamma_n+w_n+4G-2B_1^{nn})(Y_f+w_f+2M_4^{fm})-
(w_f-2B_4^{fn})(w_n+P-L_1).\label{clodiag}
\end{align}
The other notation are the same as in (\ref{obozn}). Equations
(\ref{clodiag}) describe similar interference structures as
discussed in the previous sections, but account for the effects of
driving fields on incoherent excitation. For  $E_1$ turned off,
these equations describing LICS reduce to
\begin{align}
r_n&=\left(Y_fw_n+k_nw_f\right)/\Delta_c',\quad
r_f=\left(Y_nw_f+k_fw_n\right)/\Delta_c',\nonumber\\
\Delta_c'&=(Y_f+w_f)(Y_n+w_n)-k_nk_f,\\% \quad
k_n&=2\gamma_{nn}^n\gamma_{ff}^f\Re\left(b_3^{nf}/y_{nf}\right)-w_n,\quad
k_f=2\gamma_{nn}^n\gamma_{ff}^f\Re\left(b_4^{fn}/y_{nf}\right)-w_f,\label{pop10}
\end{align}
and the dissociation rate is calculated with the aid of equation
(\ref{re}). In the case of $E_3=0$, dissociation from an
incoherently populated level $n$ is described by the equation in
the common form of
\begin{align}
\dot W/2=\gamma_{nn}^n w_n/(\Gamma_n+2\gamma_{nn}^n+w_n).
\end{align}
Two-photon dissociation from the ground level, when the field
$E_3$ is off, can be analyzed with the formula
\begin{align}
\dot
W/2=\frac{\gamma_{nn}^n(w_n+2|G_{mn}|^2\widetilde\Gamma_{mn}/|y_{mn}|^2)}
{\widetilde\Gamma_n+w_n+4|G_{mn}|^2\widetilde\Gamma_{mn}/|y_{mn}|^{2}}.
\end{align}

To a some extent, LICS can be treated as dressed multiphoton
resonances accompanied  by the interference of different quantum
pathways. As follows from (\ref{nasel}) and (\ref{clodiag}), when
such resonances are not fulfilled and $|y_{nf}|,|y_{mf}|\gg
\gamma, |G_{mn}|$, all interference structures disappear. The
possibility of coherent quantum control of photodissociation based
on the derived equations is illustrated below with the aid of a
numerical model relevant to electronic-vibration-rotation terms of
the molecule Na$_2$ in the next section.
\subsection{Numerical analysis}
Many  experiments on coherent control of branching chemical
reactions have been carried out with  sodium dimers Na$_2$. In
this case, level $m$ of our model (Fig. \ref{levb}) can be
attributed to the state $X^1\Sigma_g^+ (v=0,J=45)$ for the close
configuration and $X^1\Sigma_g^+ (v=14,J=45)$  for the open one,
levels $n$ and $f$  to the states $A^1\Sigma_u^+ (v=6,J=46)$ and
$B^1\Pi_u (v=5,J=45)$ correspondingly, and the continuum states
$\varepsilon$  to the dissociation continuum Na(3s)+Na(3s). The
relevant relaxation constants are: $\Gamma_m=2\cdot 10^7
\,c^{-1}$, $\Gamma_n= \Gamma_f=1.2\cdot 10^8\, c^{-1} $. The
relaxation rates for coherence (off-diagonal elements of the
density matrix) are estimated as half of the sum of the
corresponding rates for the diagonal elements.

\subsubsection{Coherent control of populations and dissociation
in folded schemes}

Numerical simulation for a closed-type scheme for the
quasi-stationary case shows the possibilities of manipulation of
continuum resonances through the effects of the interference
caused by a strong laser field. Due to the action of strong
radiation at a discrete transition, a split of level $n$ appeares.
Two LICS in the dissociation continuum can be detected by
variation of the detunings $\Omega_{nf}$ through the change of
frequency of the laser field at the adjacent transition to the
continuum $({f\varepsilon})$.\\

\begin{figure}[!h]\begin{center}
{\bf (a)}\hspace{60mm} {\bf (b)}\\
\includegraphics[width=0.45\textwidth]{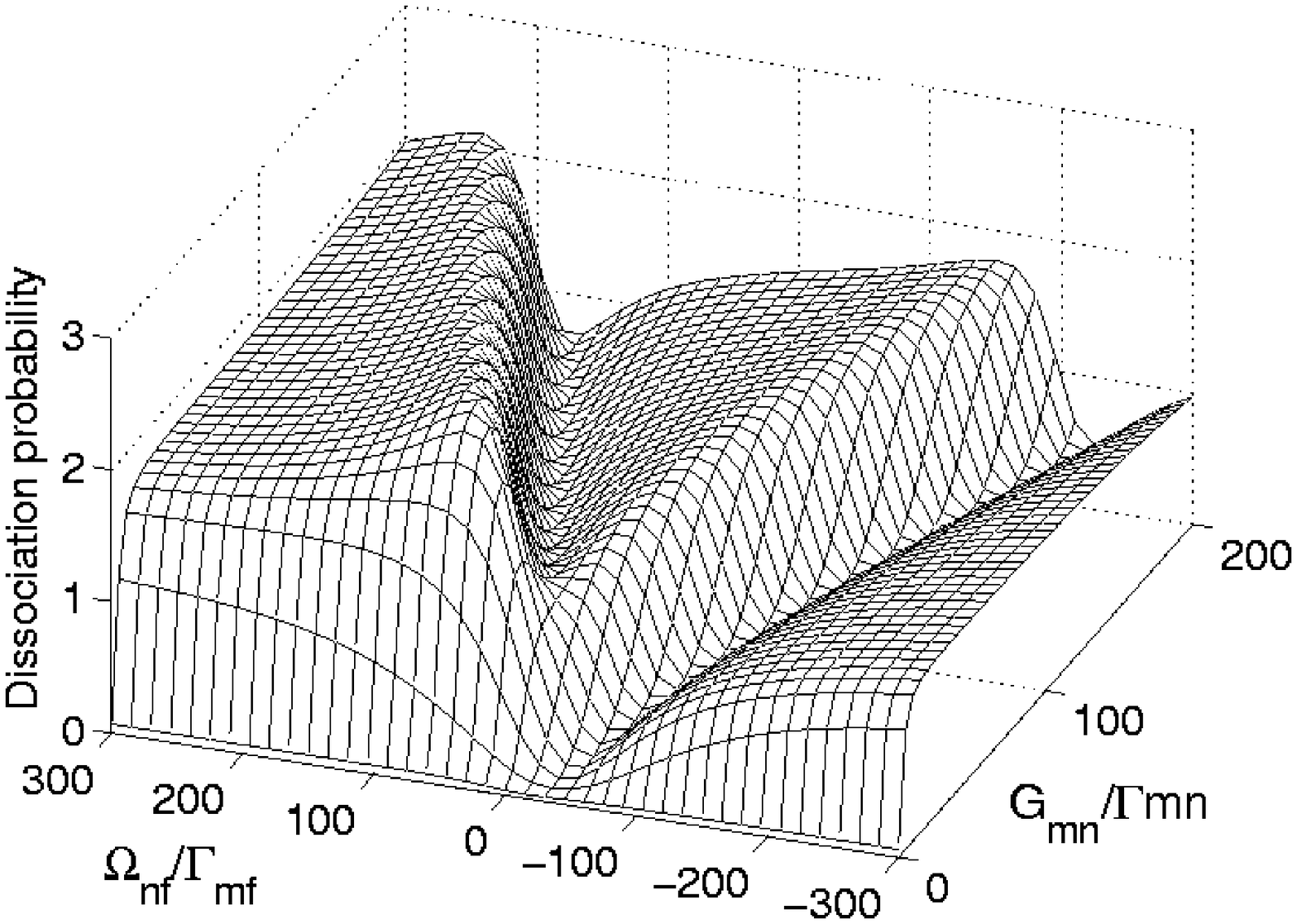}\hspace{5mm}
\includegraphics[width=0.45\textwidth]{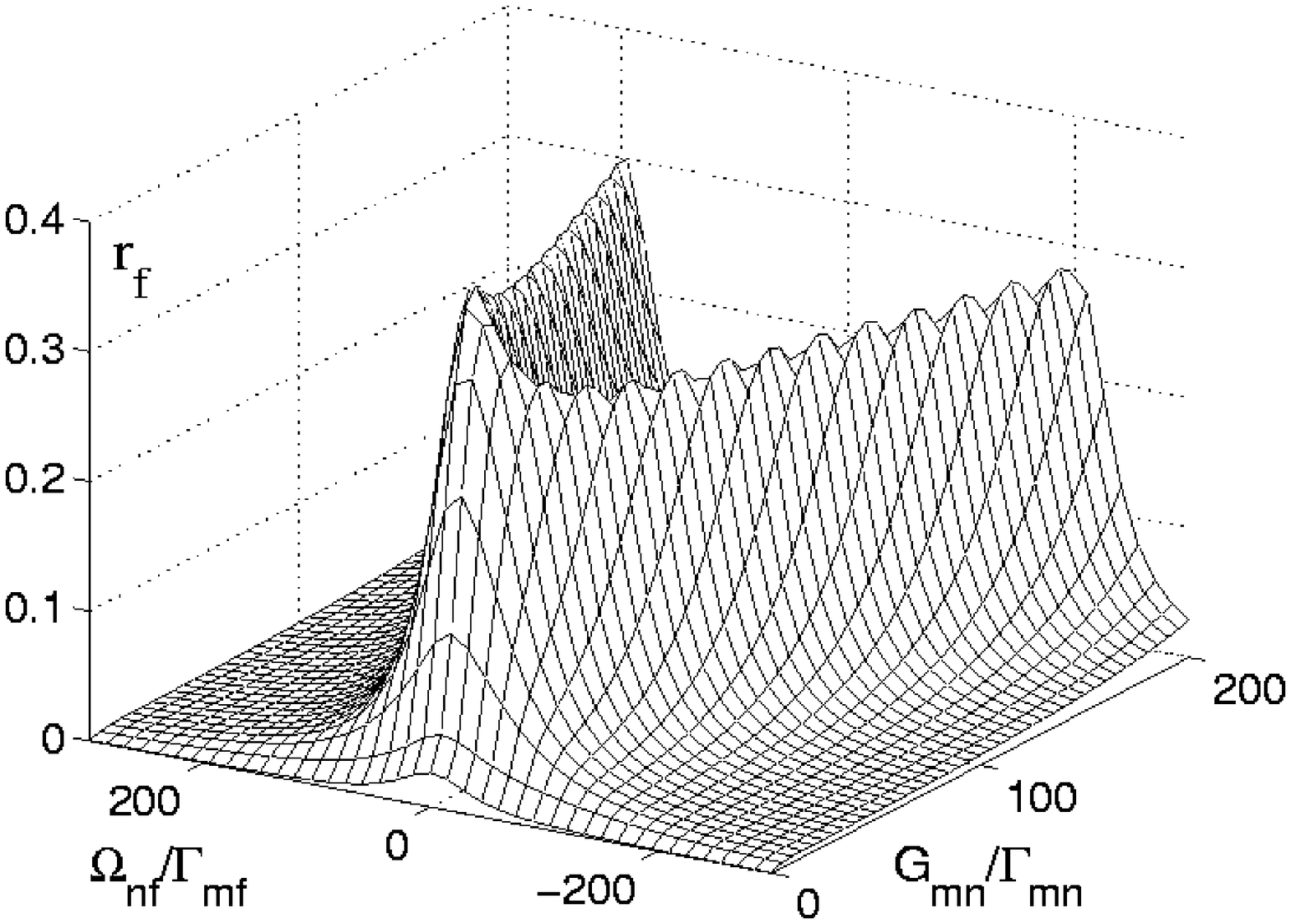}
\end{center}
\caption[recon]{\label{cl1} Dissociation probability {\bf (a)} and
population of level $f$ {\bf (b)} as functions of
$\Omega_{nf}/\Gamma_{mf}$ and $G_{mn}/\Gamma_{mn}$. Here,
$\gamma_{nn}/\Gamma_{mn}=3$, $\gamma_{ff}/\Gamma_{mf}=3$,
$q_{nn}=0.2$, $q_{ff}=-0.5$, $q_{nf}=10$, $w_{n}/\Gamma_n=0.1$,
$w_{f}=0$, $\Omega_{mn}=0$.}
\end{figure}

Figure \ref{cl1} shows that the dissociation rate can be
considerably eliminated or alteratively enhanced by varying either
the intensity of the splitting field or the detuning
$\Omega_{nf}$. The spectral lineshape  at constant field
intensities has a well-known asymmetrical Fano profile. The plot
{\bf(b)} shows opportunities  for manipulating the population of
level $f$.

\begin{figure}[!h]
\begin{center}
{\bf(a)}\hspace{60mm} {\bf (b)}\\
\includegraphics[width=0.4\textwidth]{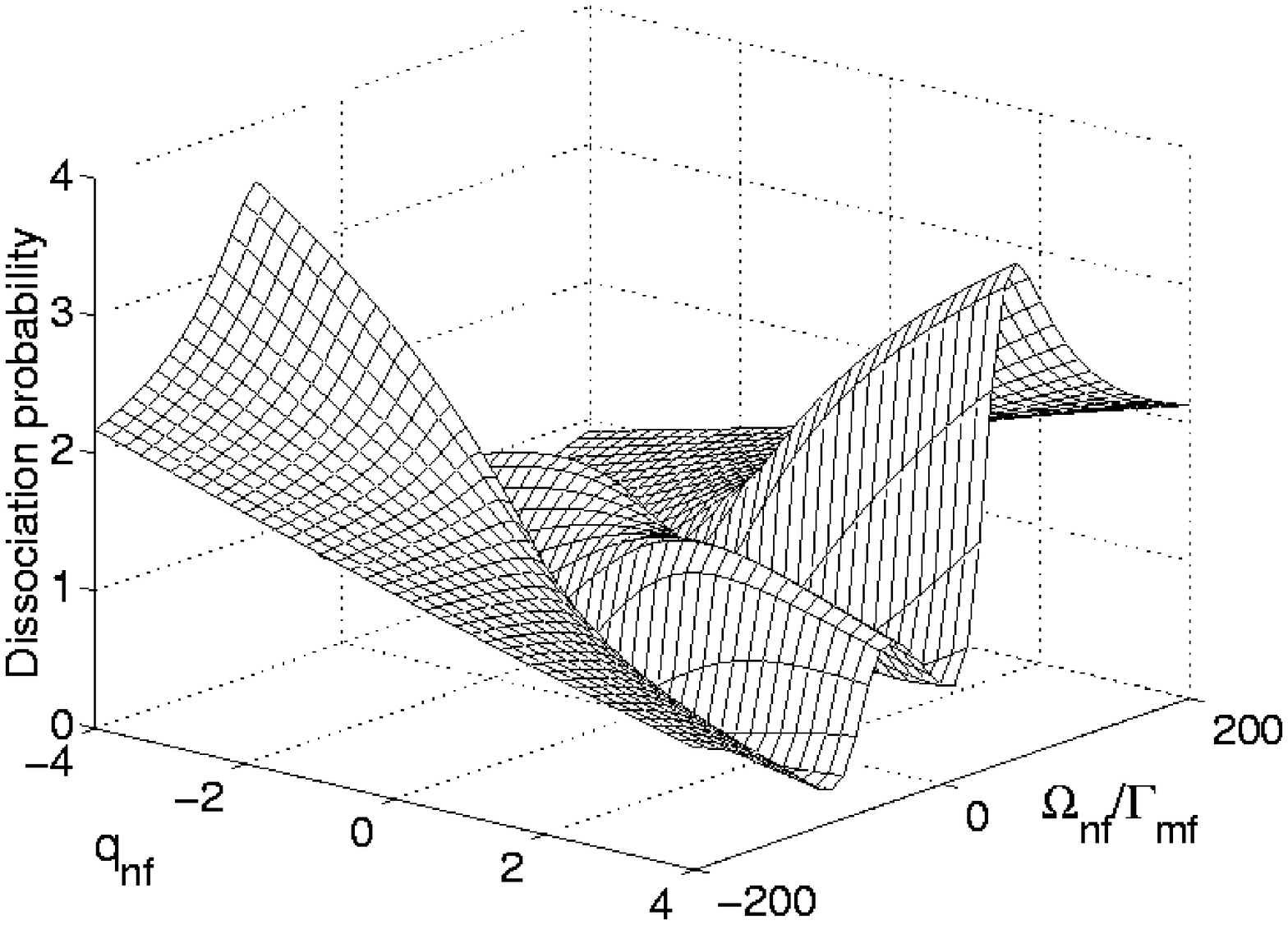}\hspace{5mm}
\includegraphics[width=0.4\textwidth]{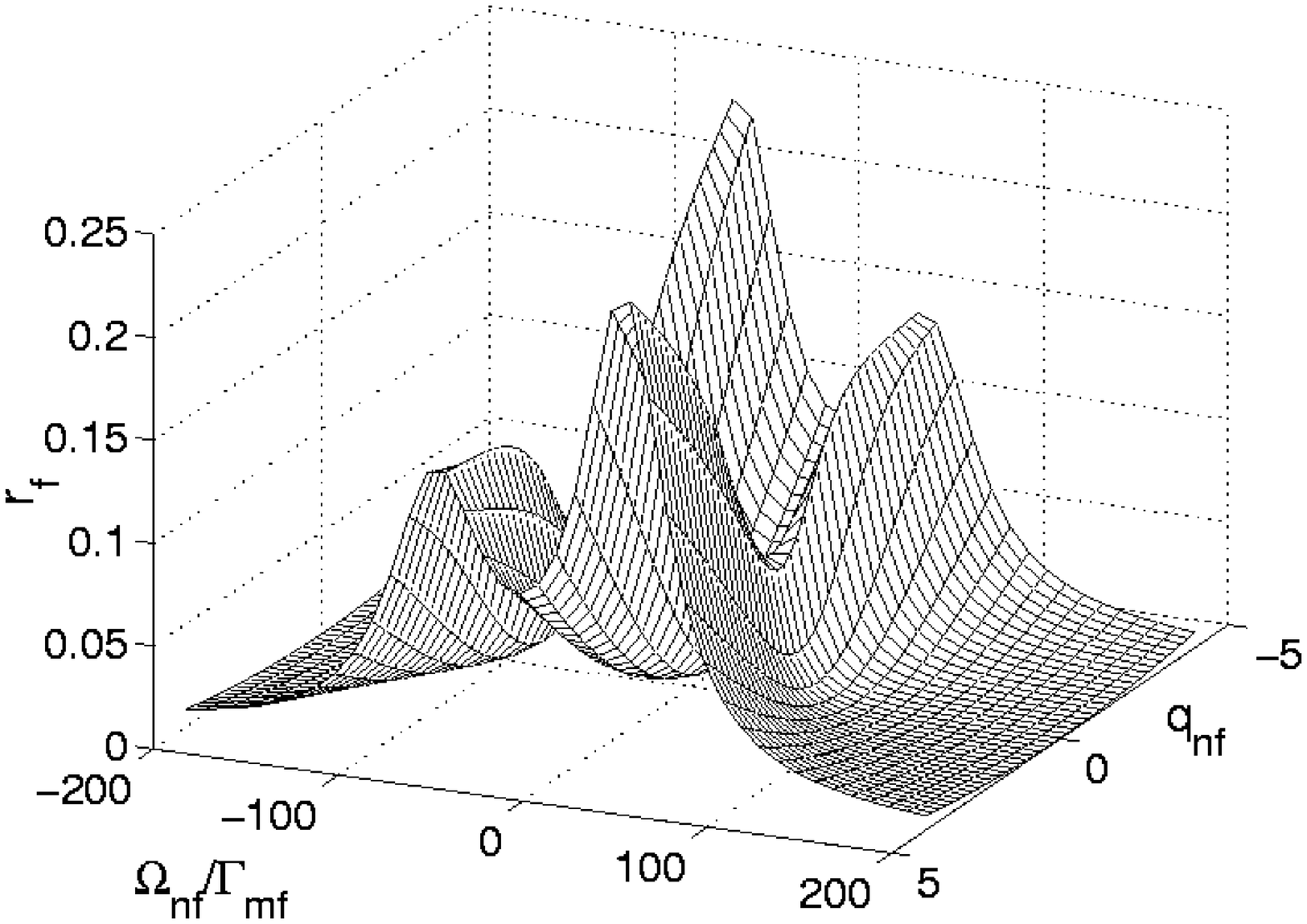}
\end{center}
\caption[recon]{ Dissociation probability {\bf (a)} and population
of level $f$ {\bf (b)} as functions of $\Omega_{nf}/\Gamma_{mf}$
and of the Fano parameter $q_{nf}$. Here, $G_{mn}/\Gamma_{mn}=50$.
The other parameters are the same as in the previous
figure.}\label{cl2}
\end{figure}

The properties of LICS are essentially determined by the Fano
parameters, which indicate an asymmetry in the distribution of the
oscillator strength over the continuum, which is illustrated in
Fig. \ref{cl2}. By varying the position of the LICS within the
continuum, the intensities of the strong laser radiation  and the
one-photon detuning $\Omega_1$, one can manipulate the properties
of the laser-induced structures. Large two-photon detunings lead
to the vanishing of any interference phenomena in the continuum
and thus is equivalent to the field $E_3$ being turned off.

\newpage
\begin{figure}[!h]
\begin{center}
{\bf (a)}\hspace{60mm} {\bf (b)}\\
\includegraphics[width=0.4\textwidth]{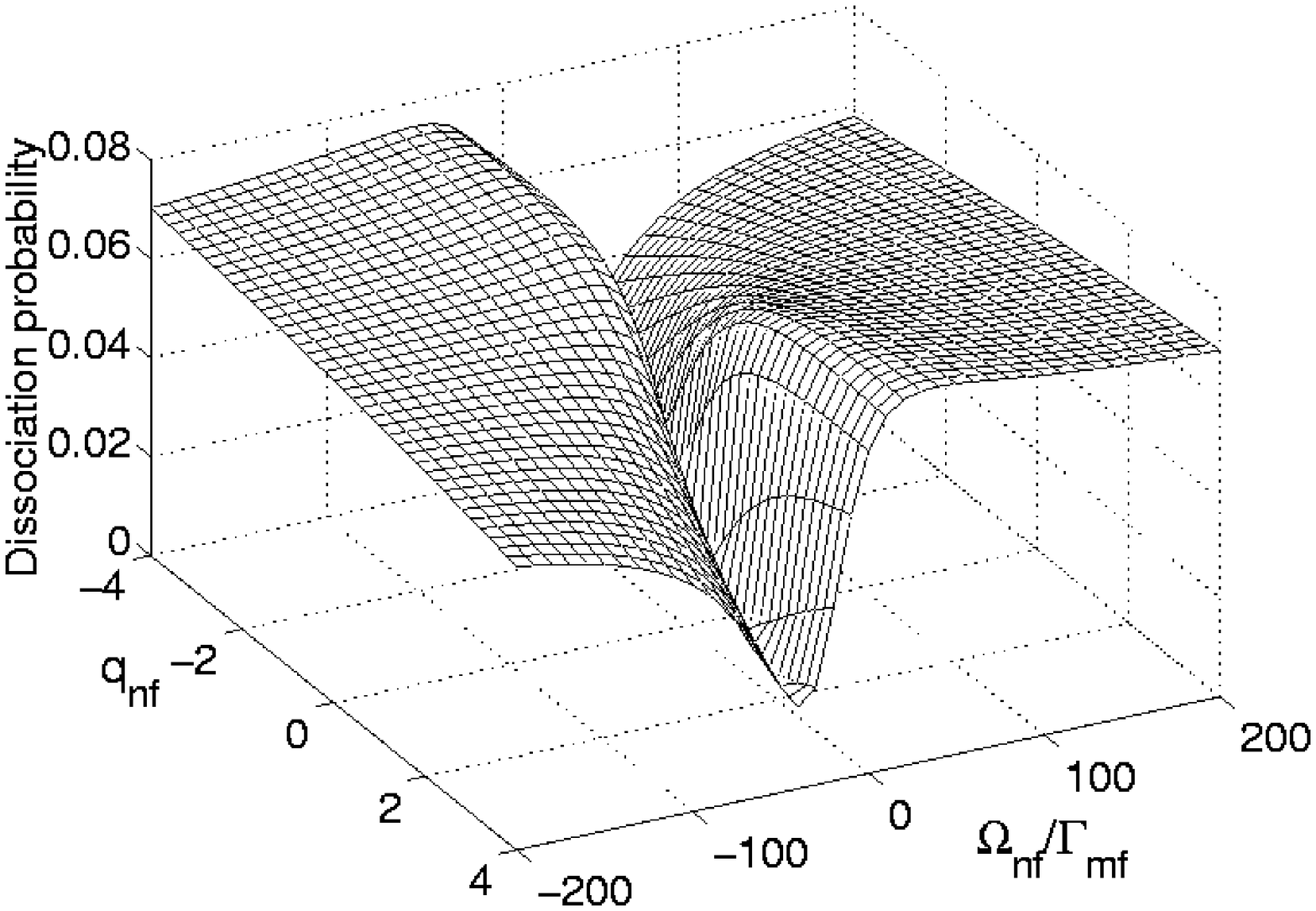}\hspace{5mm}
\includegraphics[width=0.4\textwidth]{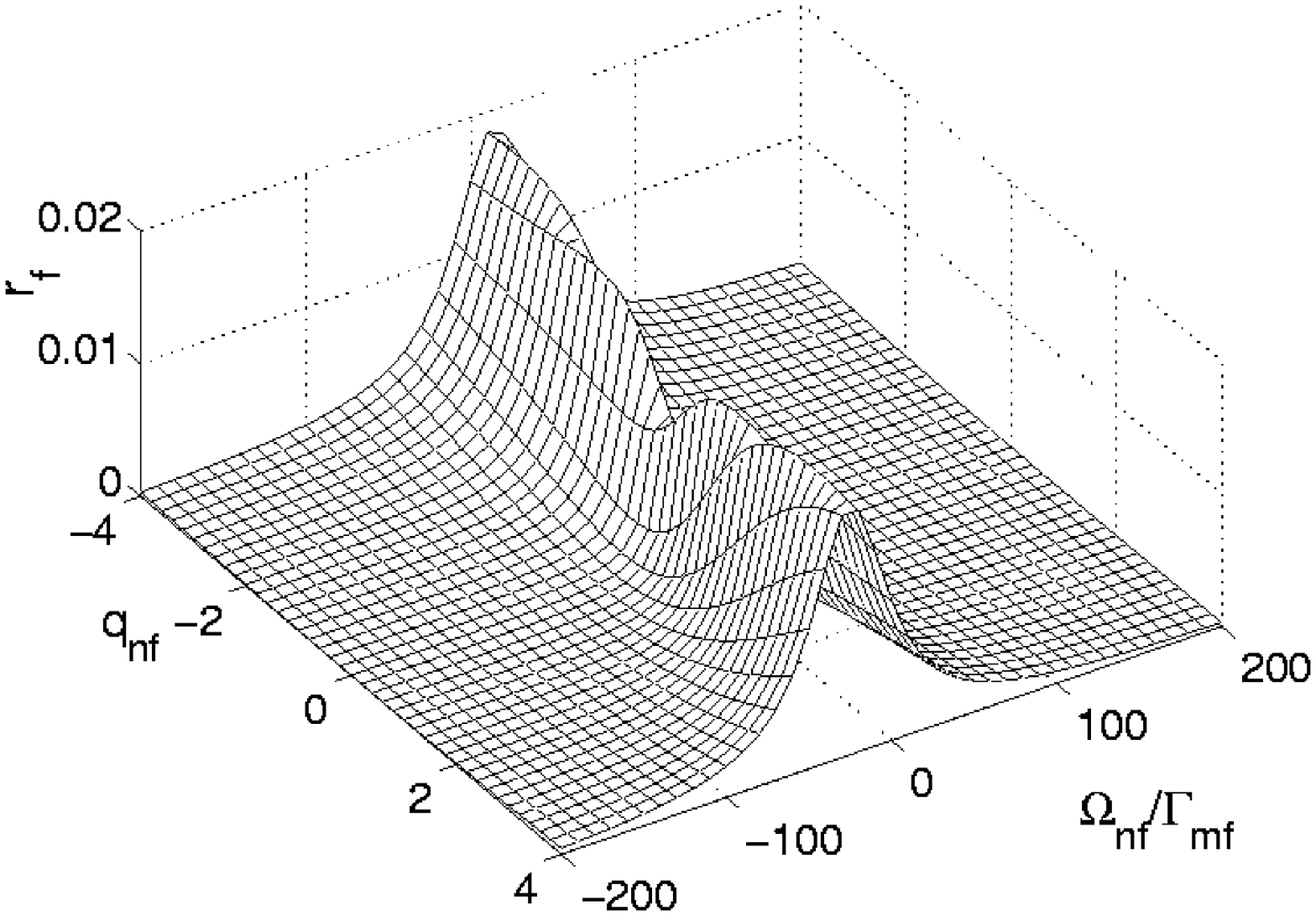}\end{center}
\caption{Dissociation probability {\bf(a)} and population of level
$f$ {\bf (b)} as functions of $\Omega_{nf}/\Gamma_{mf}$ and of the
Fano parameter $q_{nf}$. Here, $G_{mn}=0$. The other parameters
are the same as in the previous figure.}\label{cl3}
\end{figure}

The above obtained results  show that dissociation and population
transfer through the continuum can be controlled by the field
coupled to the discrete transition. Figure \ref{cl3} displays a
very trivial behavior for  $E_1$ turned off. Other opportunities
of manipulating two-photon dissociation and populations transfer
are associated with the control field coupling discrete and
continuous states.

In the next subsection, we investigate how such behavior may
exhibit itself in difference-frequency four-wave mixing and
nonlinear-optical generation of frequency-tunable radiation
through the continuous states in folded schemes.

\subsubsection{Coherent control of generated radiation in folded
schemes}

Here we consider the case where the fundamental $E_1$ and
generated $E_S$ fields  are weak. Then the  solution of the
density matrix equations  (Section \ref{qsc}) for a closed-type
scheme takes the form
\begin{align}
    &r_{mn}=iG_{mn}y_{mf}/(y_{mn}y_{mf}-b_1^{mm}),\nonumber\\
    &r_{mf}=-iG_{mn}\gamma_{nf}^m(1-iq_{nf})/(y_{mn}y_{mf}-b_1^{mm}).
\end{align}
Similarly, for the case  where the field $E_S$ is the probe
radiation and $E_1$ is the generated one,  we obtain
\begin{align}
    &R_{mf}=iG_{mf}y_{mn}/(y_{mn}y_{mf}-b_1^{mm}),\nonumber\\
    &R_{mn}=iG_{mf}\gamma_{fn}^{m}(1-iq_{fn})/(y_{mn}y_{mf}-b_1^{mm}).
\end{align}
Then the wave equations for weak fields, coupled by the four-wave
mixing process $\omega _{S }\leftrightarrow\omega _{1}-\omega
_{2}+\omega _{3}$,  have the form
\begin{align}
dE_{S }(z)/dz&={\rm i}2\pi k_{S}'\chi
^{(3)}_{S}E_{2}^{*}E_{3}E_{1}(z)\exp ({\rm i}\Delta kz),
\nonumber\\
dE_{1}(z)/dz&={\rm i}2\pi k_{1}'\chi ^{(3)}_{1} E_{2}E^{*}_{3}E_{S
}(z)\exp (-{\rm i}\Delta kz)\label{volsisf},
\end{align}
where $\Delta k=k_{S }-k_{1}+k_{2}-k_{3}$.
\begin{figure}[!h]\begin{center}
{\bf (a)}\hspace{70mm}{\bf (b)}\\
\includegraphics[width=0.45\textwidth]{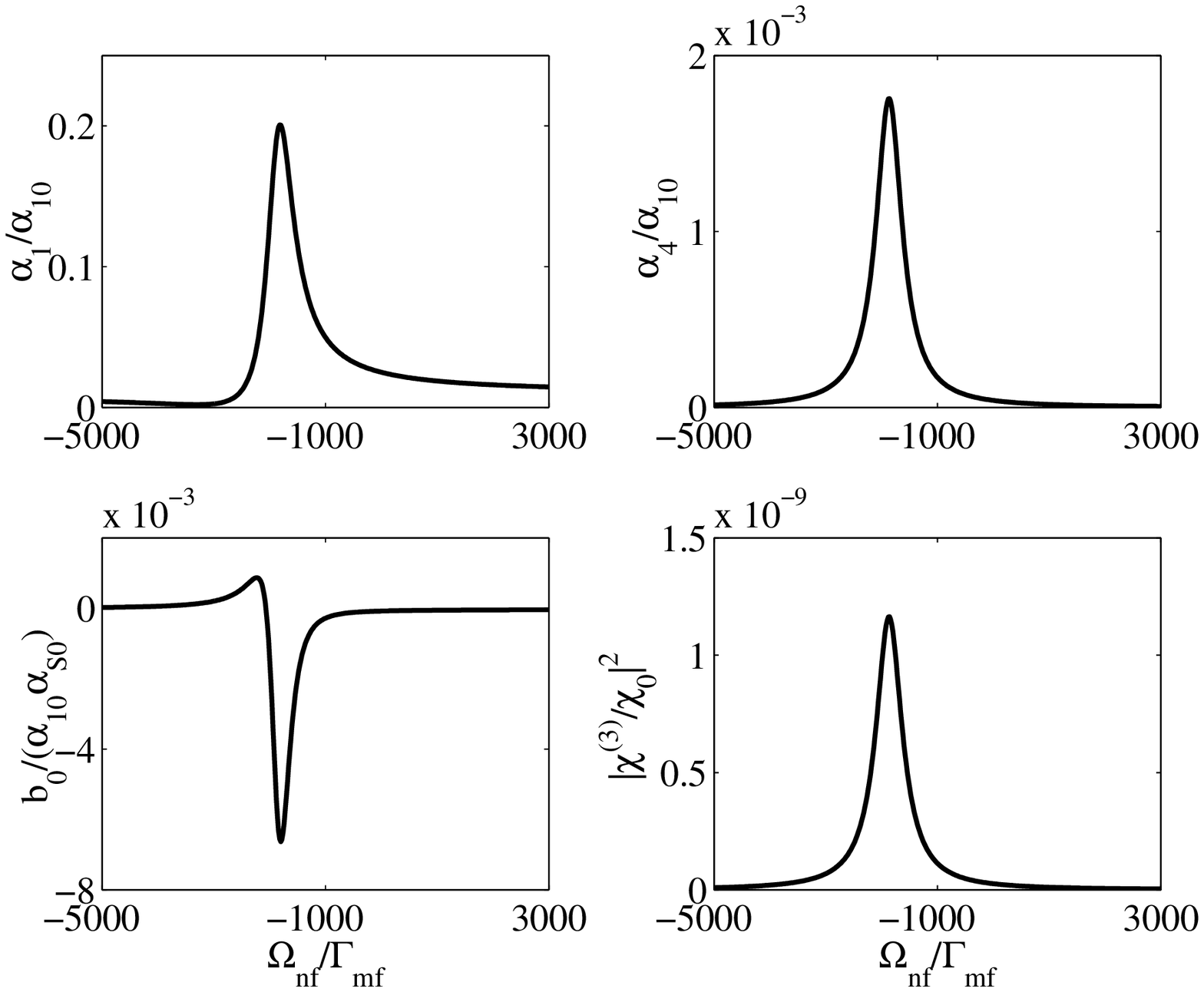}\hspace{5mm}
\includegraphics[width=0.45\textwidth]{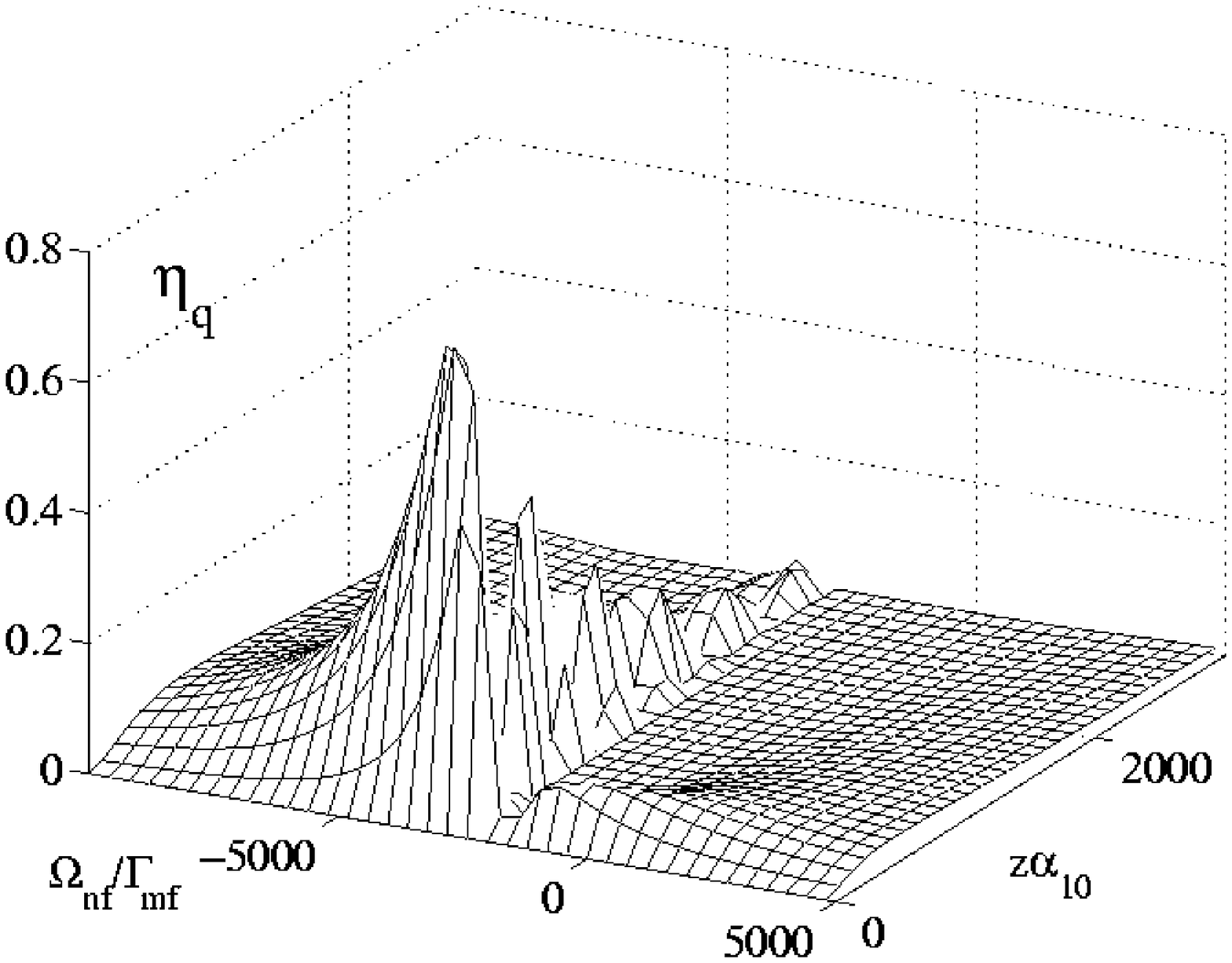}\end{center}
\caption{Dependence of the absorption coefficients $\alpha_1 /
\alpha_{10}$ and $\alpha_S/\alpha_{10}$, of the conversion rate
$b/(\alpha_{10}\alpha_{40})$, and of the squared modulus of the
nonlinear susceptibility $|\overline\chi^{(3)}|^2$, on the
detuning $\Omega_{nf}$ {\bf (a)}, and dependence of the quantum
conversion efficiency $\eta_{\rm q}$ on the optical thickness and
on $\Omega_{nf}$ {\bf (b)}. Here, $C=0.5$, $g_{n}=100$,
$g_{f}=6000$, $\Omega_1/\Gamma_{mn}=0$, $q_{ff}=-0.5$,
$q_{nn}=0.2$, $q_{nf}=0$, {\bf (a,b)}.} \label{gener1}
\end{figure}
 Analogously to (\ref{volsisl}), the solution of equations (\ref{volsisf}) can be
presented like  (\ref{qen}) with $\overline\eta_{{\rm q}0}=
\eta_{{\rm q}0}^0 |\overline \chi^{(3)}|^2 g_{n}g_{f}$, where
$g_n=\gamma_{nn}/\Gamma_{mn}$, $g_f=\gamma_{ff}/\Gamma_{mf}$. With
the aid of  the approximate expressions
\begin{eqnarray}
\alpha_{10}=4\pi {\omega_1}{|d_{mn}|^2/ c \hbar\Gamma_{mn}},\>
\alpha_{S0}=4\pi {\omega_S}|d_{mf}|^2/c \hbar\Gamma_{mf},
\nonumber\\
|\chi_0|^2=(\pi/8\hbar^2)^2(1+q_{nf}^2)|d_{mn}d_{n\varepsilon}d_{\varepsilon
f}d_{fm}|^2 (\Gamma_{mn}\Gamma_{mf})^{-2},
\end{eqnarray}
we obtain that $\eta_{{\rm q}0}^0=(\pi/2)^2(1+q_{nf}^2)$. Other
notations are the same as in Section \ref{gl}.
\begin{figure}[!h]\begin{center}
{\bf (a)}\hspace{70mm}{\bf (b)}\\
\includegraphics[width=0.45\textwidth]{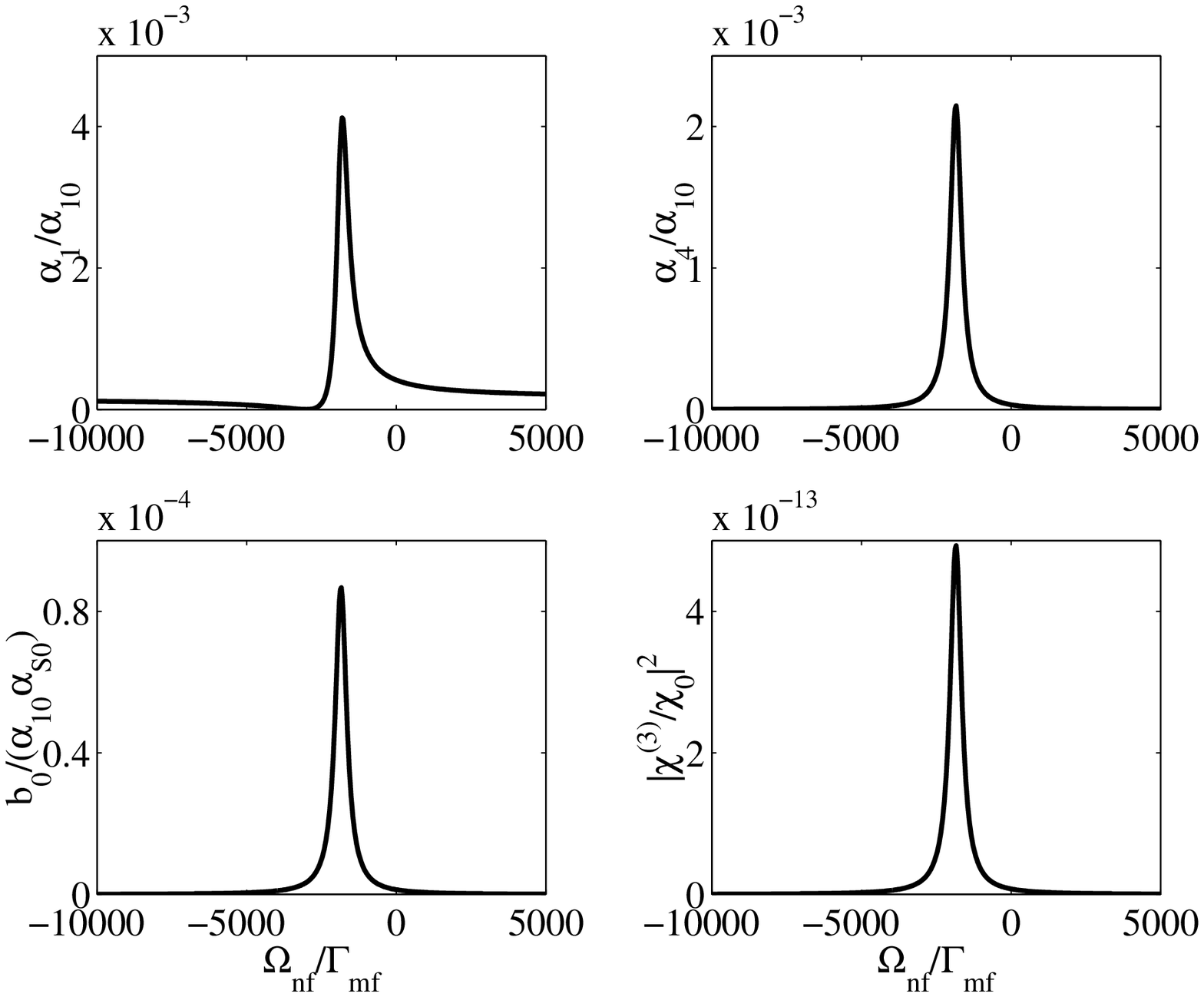}\hspace{5mm}
\includegraphics[width=0.45\textwidth]{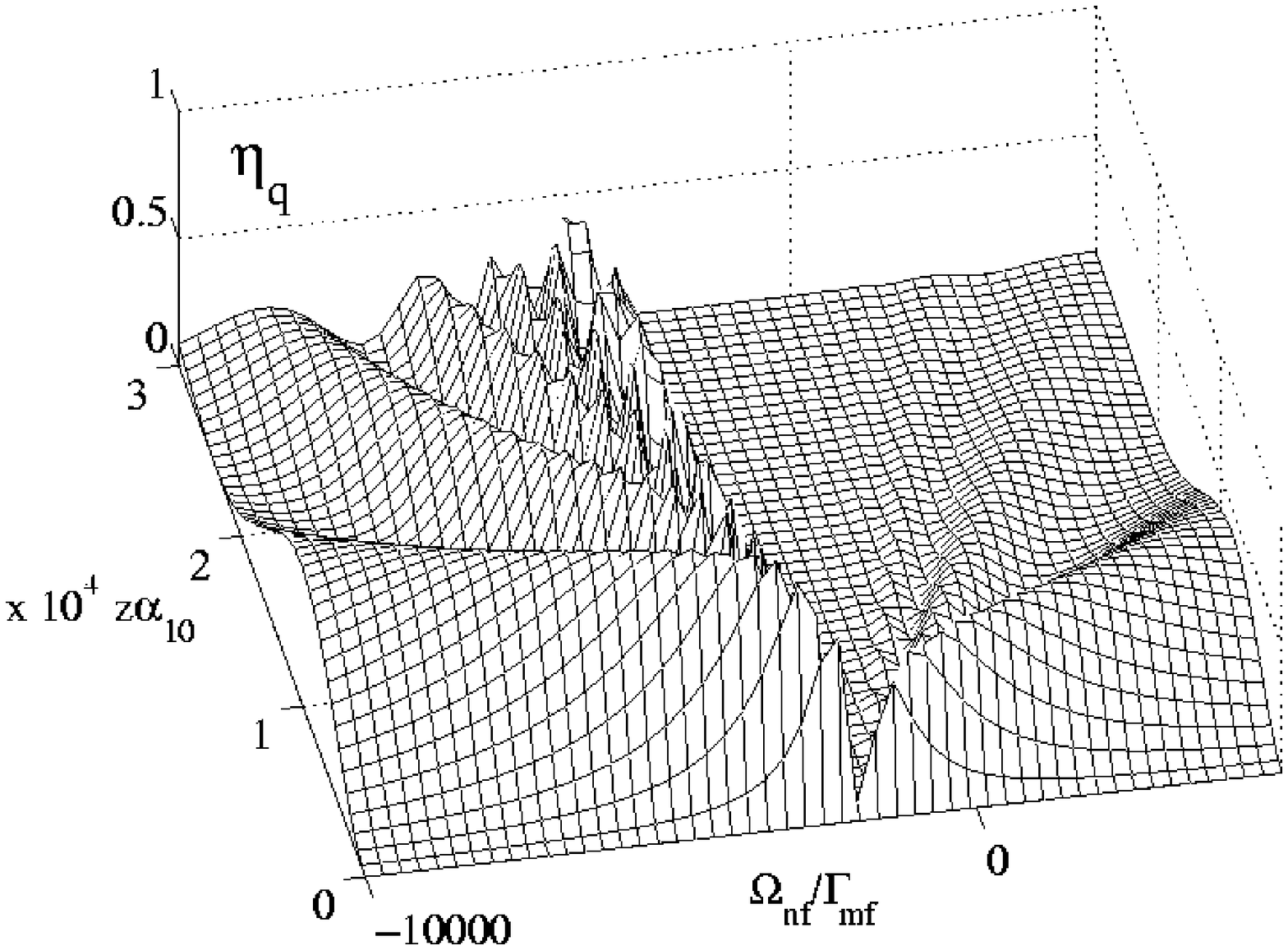}\\
{\bf (c)}\\
\includegraphics[width=0.5\textwidth]{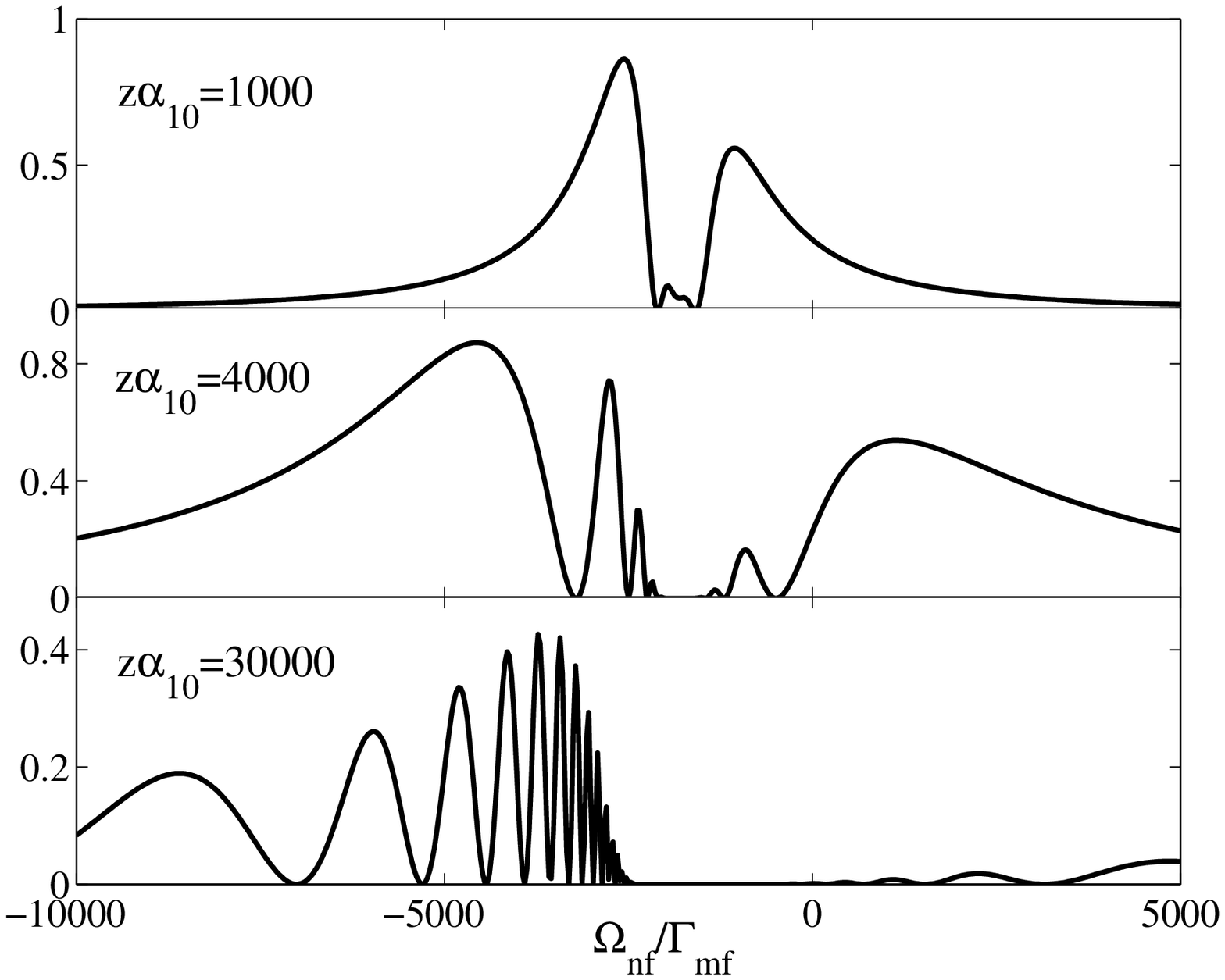}\end{center}
\caption{Dependence of the absorption coefficients $\alpha_1 /
\alpha_{10}$ and $\alpha_S/\alpha_{10}$, of the conversion rate
$b/(\alpha_{10}\alpha_{40})$, and of the squared modulus of the
nonlinear susceptibility $|\overline\chi^{(3)}|^2$ on the detuning
$\Omega_{nf}$ {\bf (a)}. Dependence of the quantum conversion
efficiency  $\eta_{\rm q}$ on $\Omega_{nf}$ for
$z\alpha_{10}=10^3$; $z\alpha_{10}=4\cdot 10^3$;
$z\alpha_{10}=3\cdot 10^4$ {\bf (b)}. Dependence of the quantum
conversion efficiency $\eta_{\rm q}$ on the optical thickness and
on $\Omega_{nf}$ {\bf (c)}. Here, $g_{n}=6000$, and the other
parameters are the same as in the previous figure.} \label{gener2}
\end{figure}

\begin{figure}[!h]\begin{center}
{\bf (a)}\hspace{70mm}{\bf (b)}\\
\includegraphics[width=0.45\textwidth]{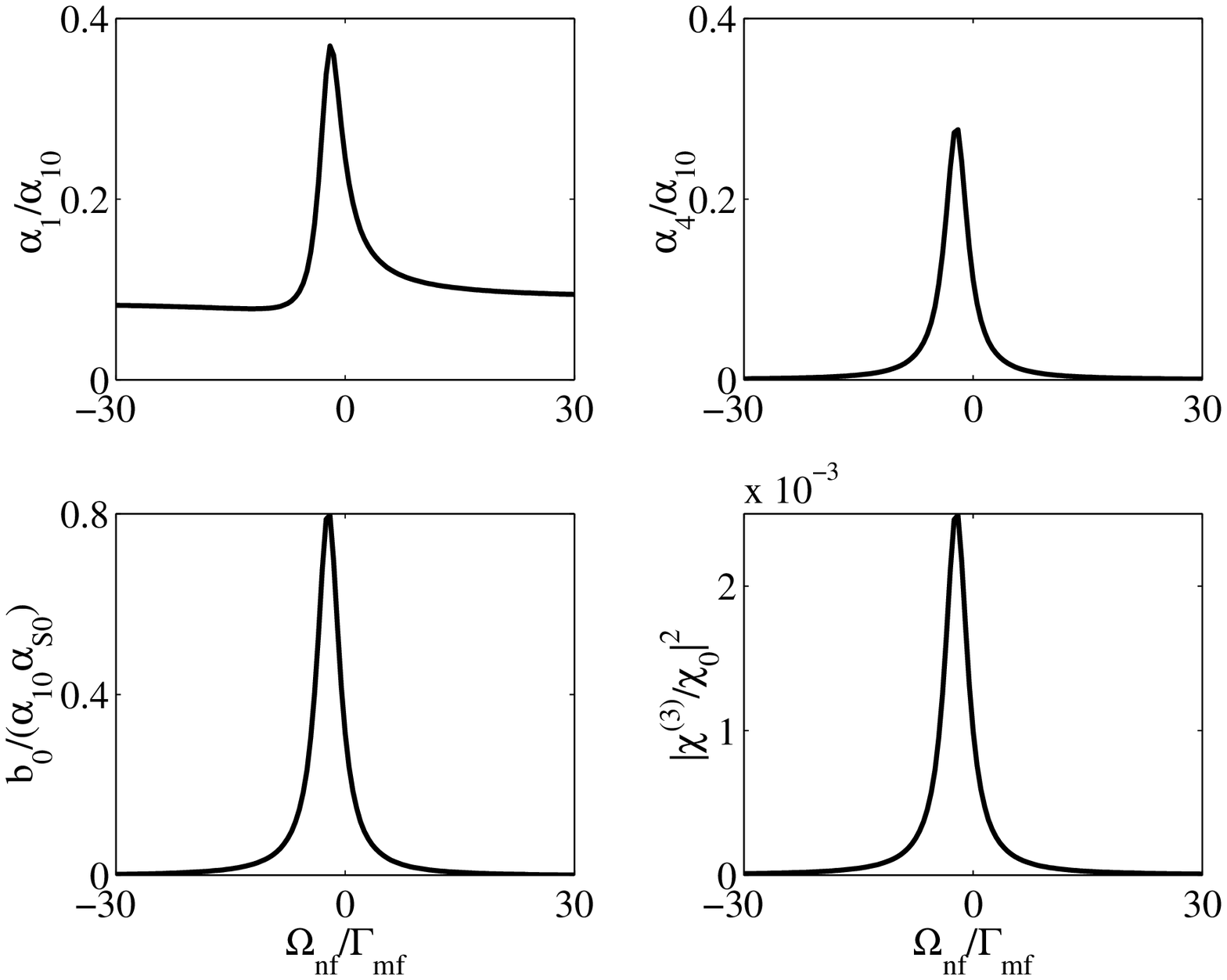}\hspace{5mm}
\includegraphics[width=0.45\textwidth]{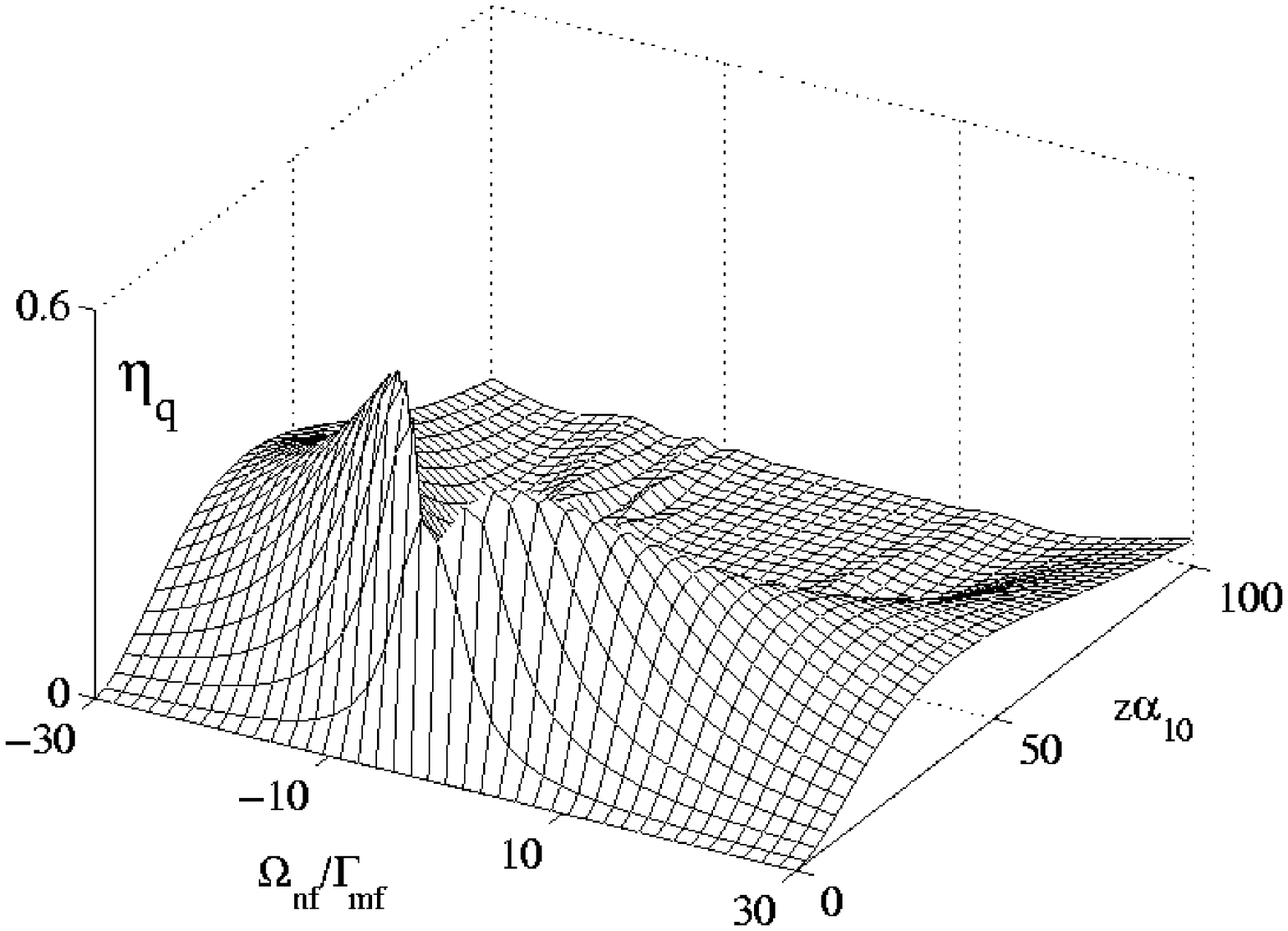}\end{center}
\caption{ Dependencies of the absorption coefficients $\alpha_1 /
\alpha_{10}$ and $\alpha_S/\alpha_{10}$, of the rate of conversion
$\overline b$, and of the square of the modulus of the nonlinear
susceptibility $|\overline\chi^{(3)}|^2$ on the detuning
$\Omega_{nf}$ {\bf (a)}. Dependence of the quantum conversion
efficiency $\eta_{\rm q}$ on the optical thickness and on
$\Omega_{nf}$ {\bf (b)}. Here, $g_{n}=10$, $g_{f}=6.5$. The other
parameters are the same as in the previous figure.} \label{gener3}
\end{figure}

\begin{figure}[!h]\begin{center}\vspace{20mm}
{\bf (a)}\hspace{70mm}{\bf (b)}\\
\includegraphics[width=0.45\textwidth]{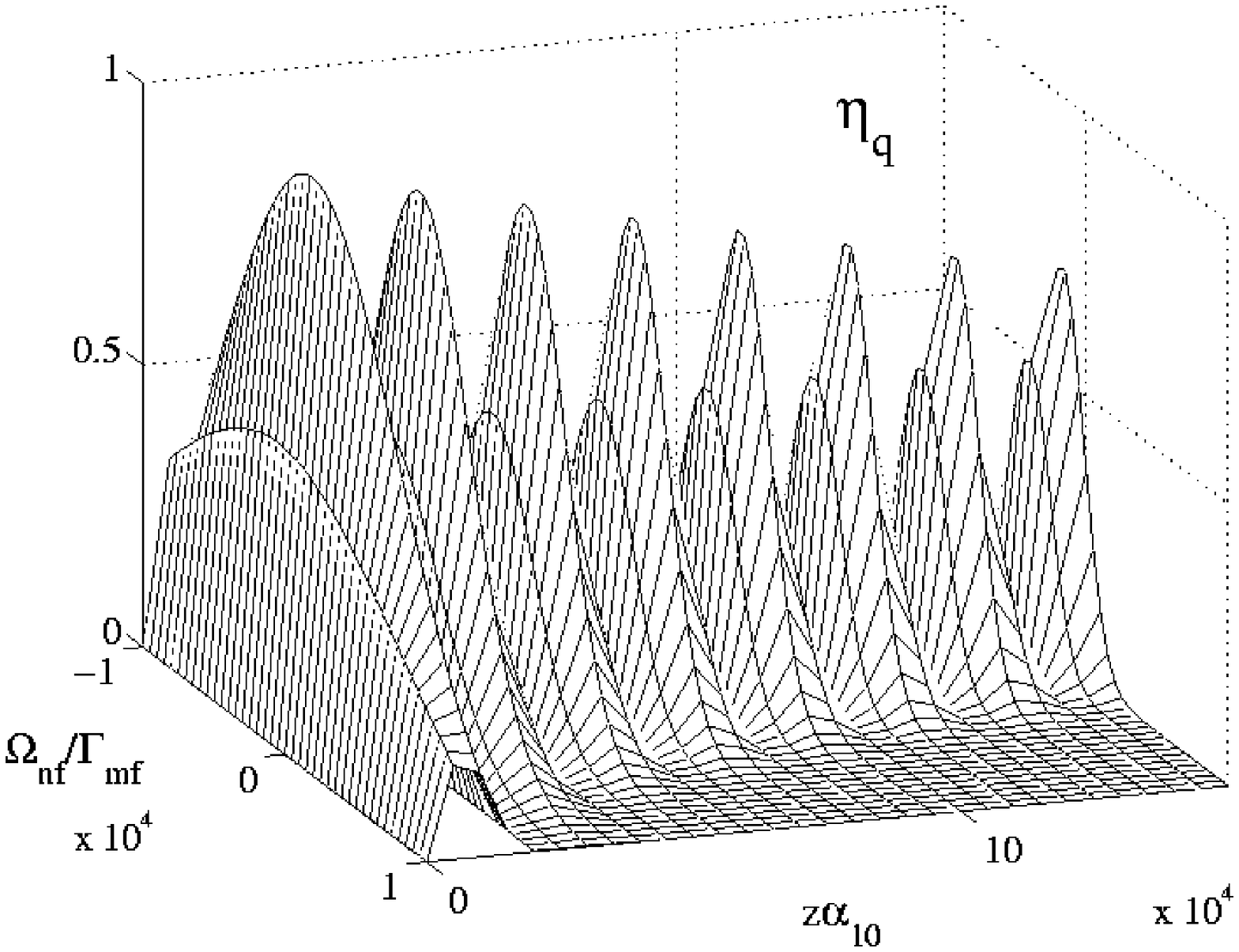}\hspace{5mm}
\includegraphics[width=0.45\textwidth]{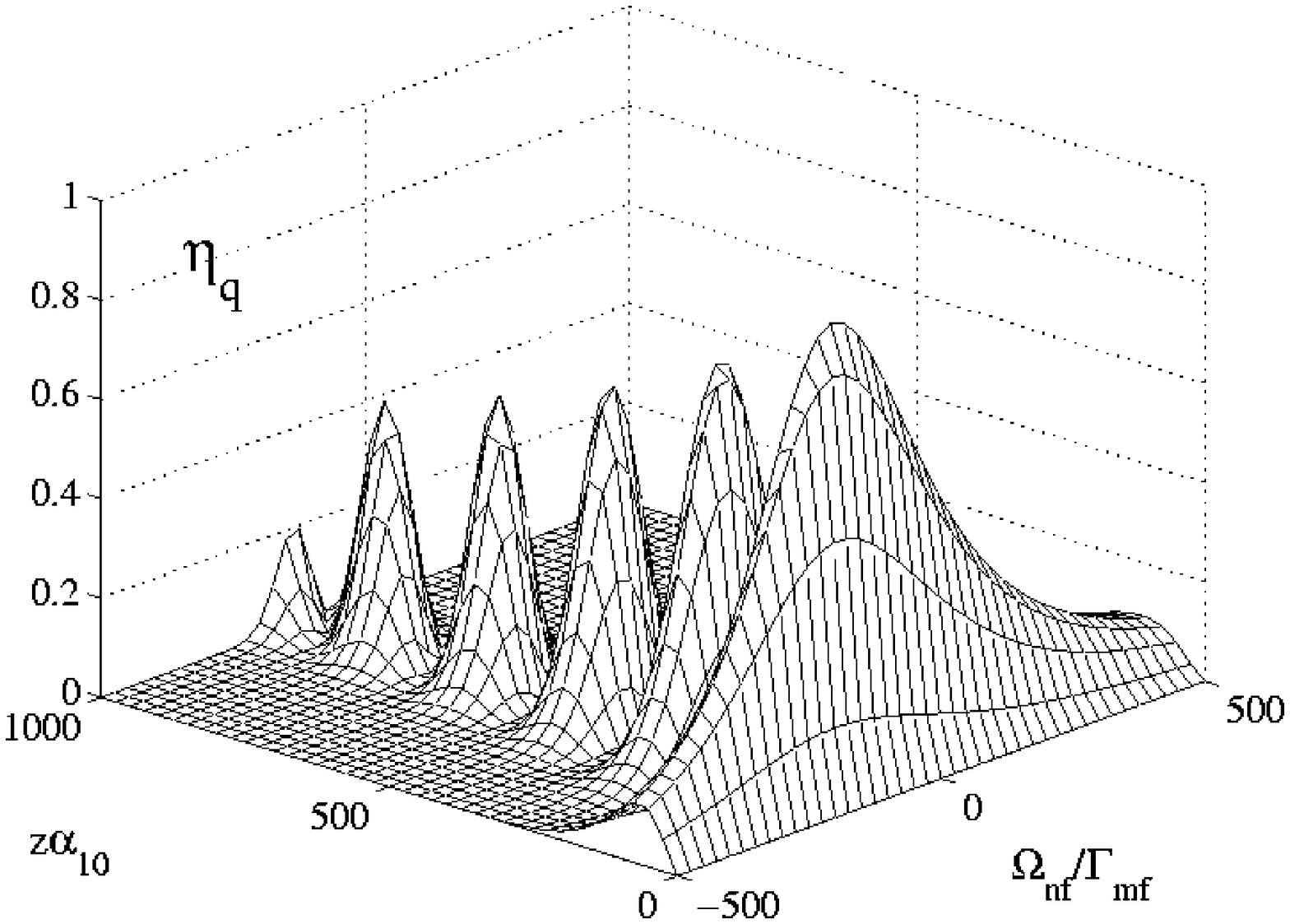}\\
{\bf (c)}\\
\includegraphics[width=0.45\textwidth]{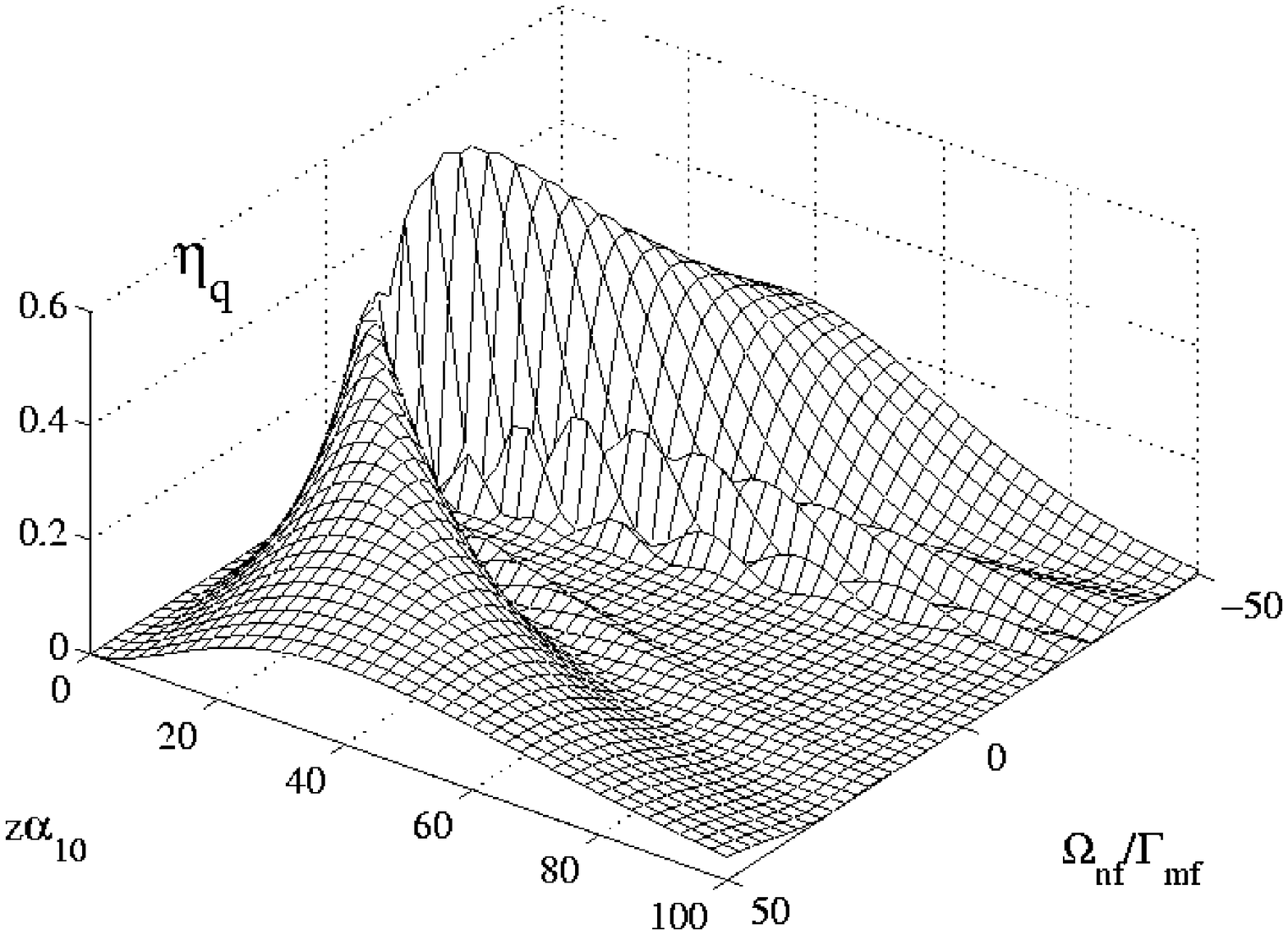}\end{center}\vspace{10mm}
\caption{Dependence of the quantum conversion efficiency
$\eta_{\rm q}$ on the optical thickness and on $\Omega_{nf}$ for
different $q_{nf}$. Here, $g_{n}=10$, $g_{f}=20$; $q_{nf}=100$
{\bf (a)}, $q_{nf}=5$ {\bf (b)}, $q_{nf}=0$ {\bf (c)}. The other
parameters are the same as in the previous figure.}
\label{gener4}\vspace{30mm}
\end{figure}
%%%%%%%%%%%%%%%%%%%%%%%%%%%%%%%%%

%%%%%%%%%%%%%%%%%%%%%%%%%%%%%%%%%%%%%
\begin{figure}[!h]\begin{center}
{\bf (a)}\hspace{70mm}{\bf (b)}\\
\includegraphics[width=0.45\textwidth]{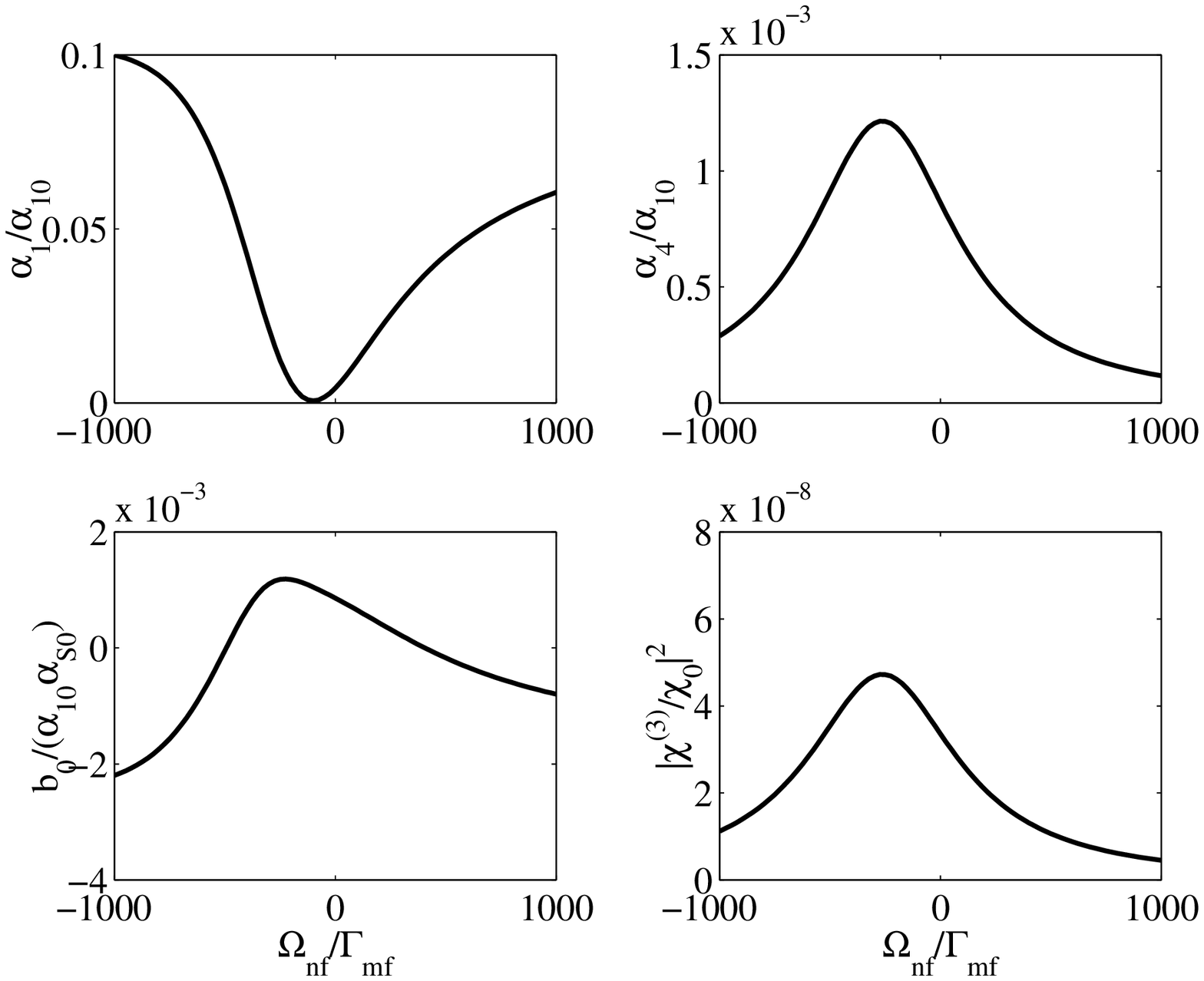}\hspace{5mm}
\includegraphics[width=0.45\textwidth]{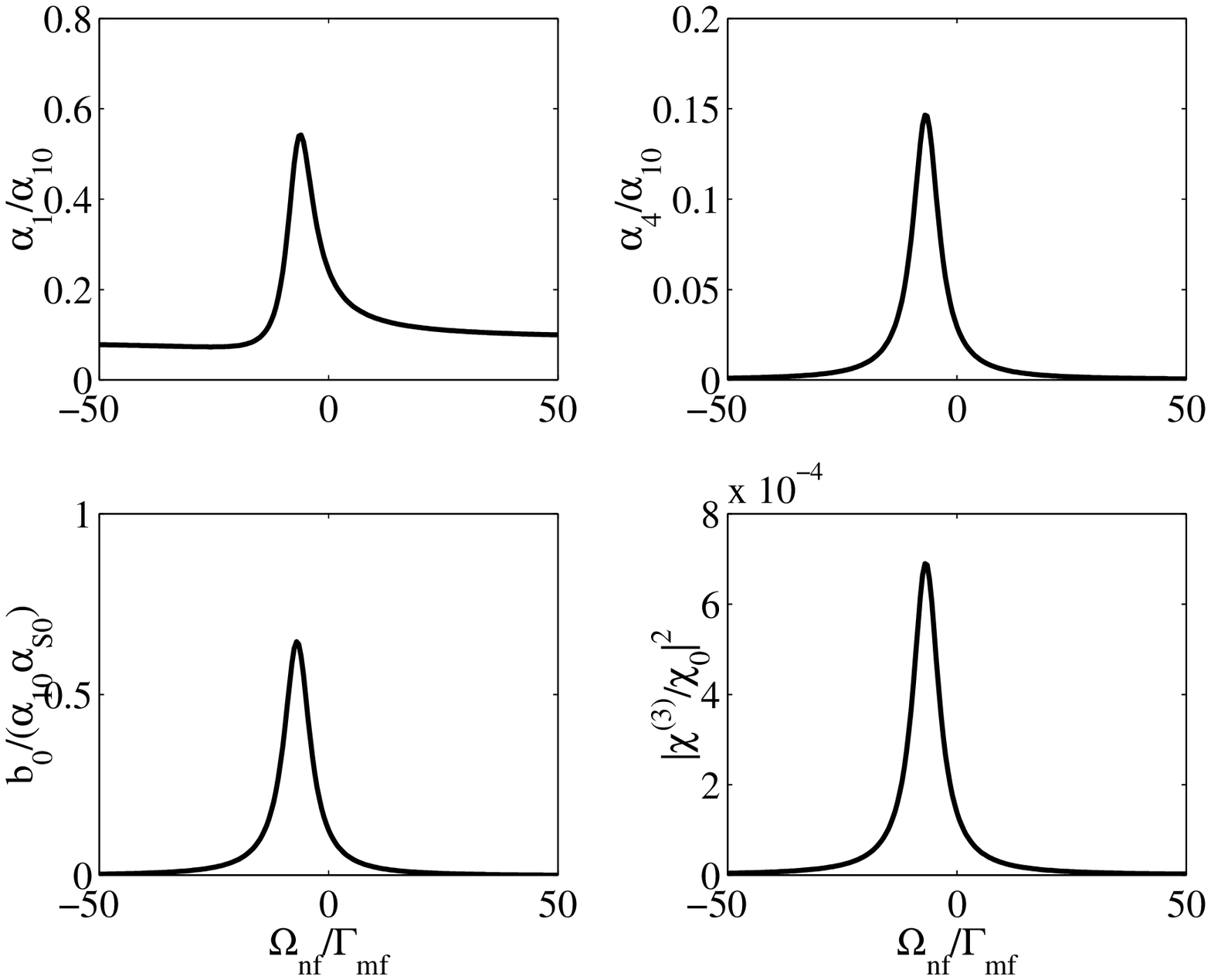}\end{center}
 \caption{Dependence of the absorption coefficients
$\alpha_1 / \alpha_{10}$ and $\alpha_S/\alpha_{10}$, of the rate
of conversion $\overline b$, and of the squared modulus of the
nonlinear susceptibility $|\overline\chi^{(3)}|^2$ on the detuning
$\Omega_{nf}$. Here, $q_{nf}=5$ {\bf (a)}, $q_{nf}=0$ {\bf (b)}.
The other parameters are equal to those in Fig. \ref{gener4}.}
\label{gener5}
\end{figure}

\begin{figure}[!h]%%6
\begin{center}
{\bf (a)}\hspace{60mm}{\bf (b)}\\
\includegraphics[width=0.45\textwidth]{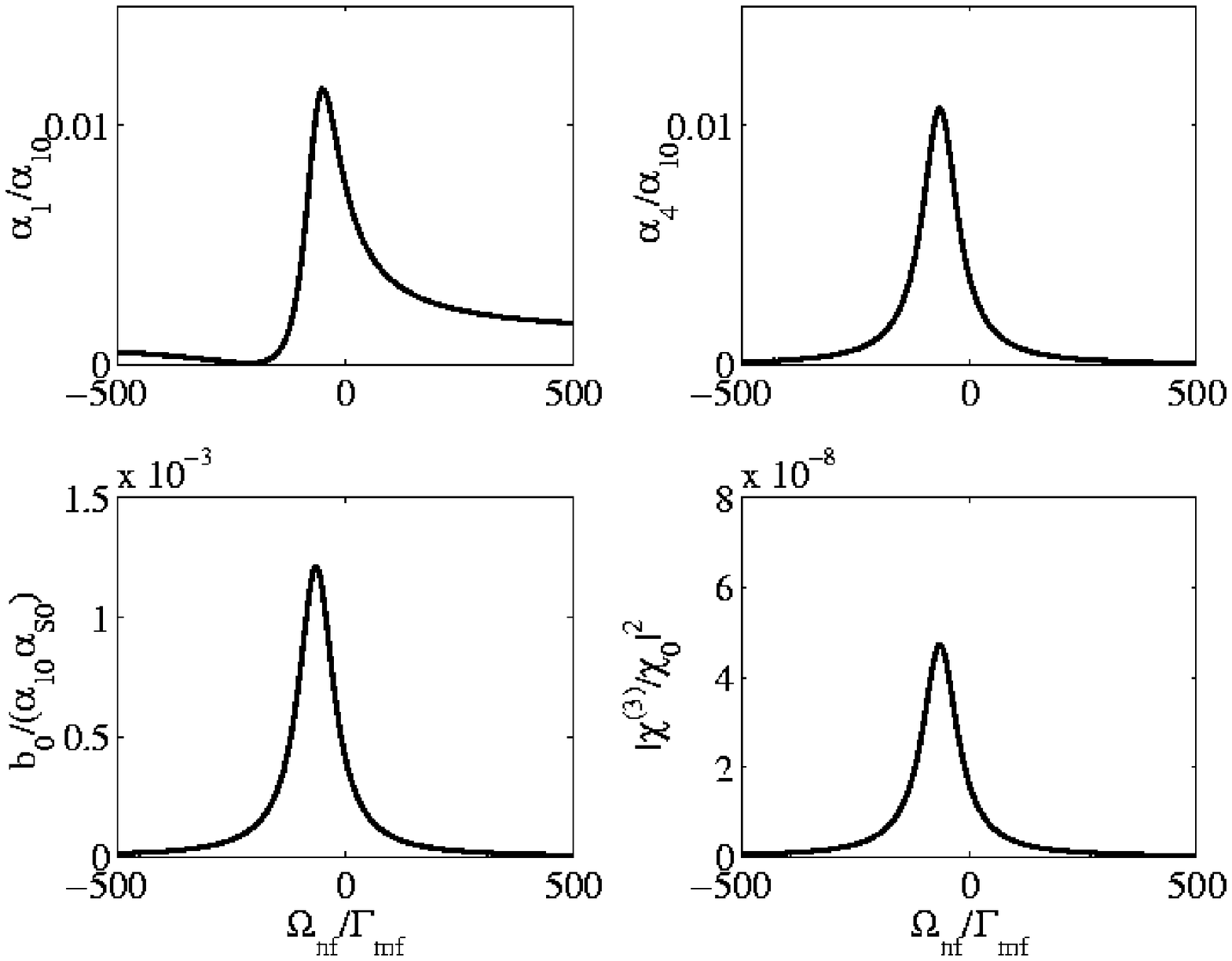}\hspace{5mm}
\includegraphics[width=0.45\textwidth]{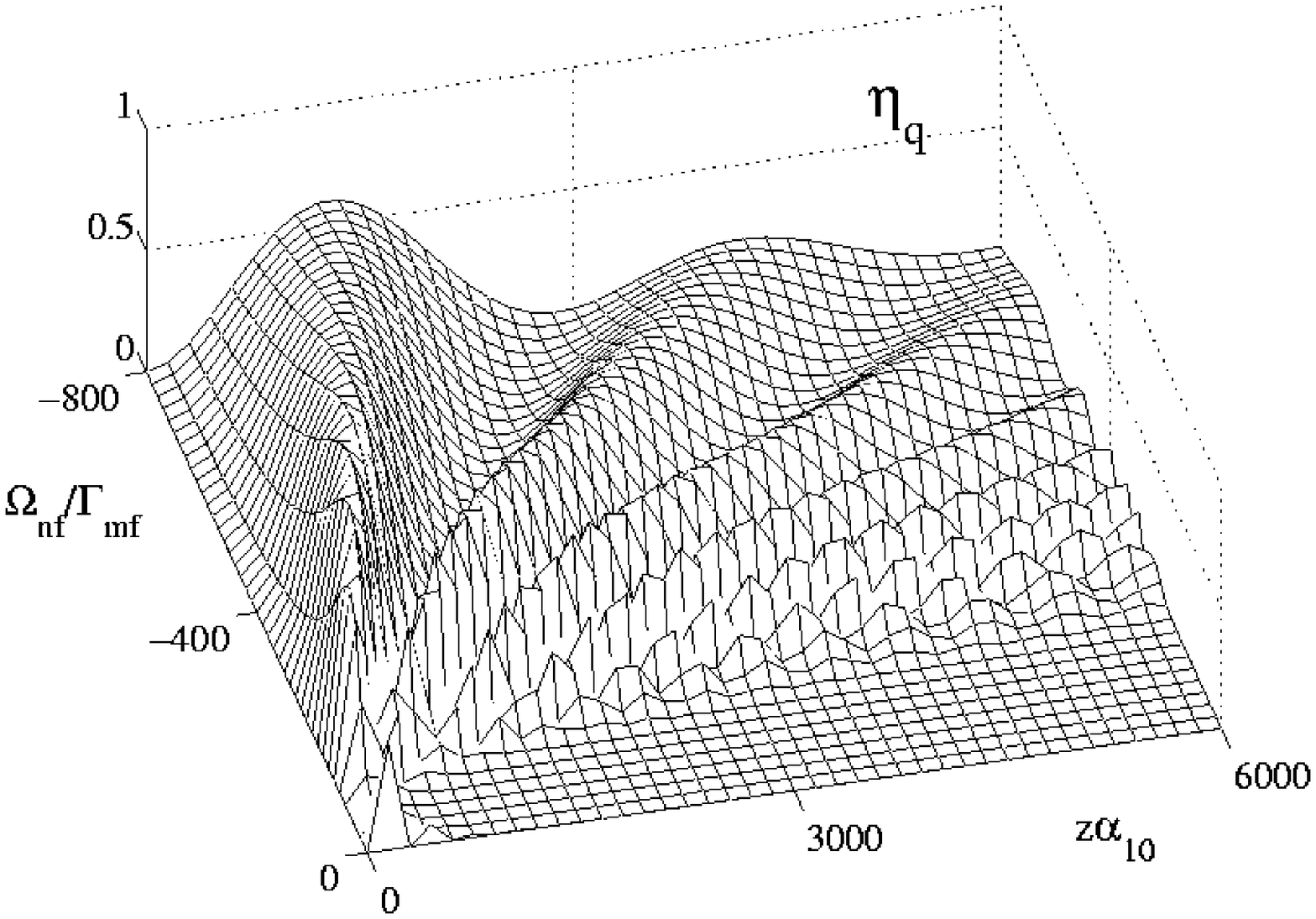}\end{center}
\caption{Dependence of the absorption indices $\alpha_1 /
\alpha_{10}$ and $\alpha_S/\alpha_{10}$, of the conversion rate
$\overline b$, and of the squared modulus of the nonlinear
susceptibility $|\overline\chi^{(3)}|^2$ on the detuning
$\Omega_{nf}$ {\bf (a)}. Dependence of the quantum conversion
efficiency  $\eta_{\rm q}$ on the optical thickness and on
$\Omega_{nf}$ [$\max(\eta_q)=0.87$] {\bf (b)}. Here $q_{nf}=5$,
$\Omega_{mn}/\Gamma_{mn}=100$. The other parameters are the same
as in the previous figure.} \label{gener6}
\end{figure}

Figure \ref{gener1} illustrates the case where the strong fields
are so large that power-broadening exceeds the half-width of the
transitions ($\Gamma_{ij}$), and also $g_f>>g_n$. Then the
generation maximum falls to the parameter  range for which
absorbtion of the fundamental radiation becomes minimum  and the
rate of conversion is positive. With that, an oscillation regime
appears. Within the interval of $\Omega_{nf}/\Gamma_{mf}$ from
$-1000$ to $3000$, the rates of absorbtion and generation are
almost equal ($b_0\approx 0$), and the oscillations reduce to only
one maximum of $\eta_q$. Figure \ref{gener2} depicts the case
where both driving parameters are equal ($g_n=g_f=6000$). Here the
maximum of $\eta_q$ also falls within the range of minimum
absorption of the fundamental radiation, and a conversion dip
corresponds to the maximum of $\alpha_1$. Interference of LICS
along with splitting of the discrete resonance give rise to
complex structures in $\eta_q$ while the waves propagate through
the nonlinear medium [see Fig. \ref{gener2}(c)]. For the driving
parameters about three order less than in the previous case, and
$g_n=g_f$,  the implementation of interference allows one to
achieve four orders of increase of the conversion rate $b_0$ (Fig.
\ref{gener3}). The maximum in $\eta_q$ becomes somehow smaller but
is reached at a much shorter length of the nonlinear medium. With
greater Fano parameters $q_{nf}$, the potential enhancement of the
output of generation through manipulating the interplay of LICS
becomes greater. As shown in Fig. \ref{gener4}, $\eta_q\rightarrow
1$ with $q_{nf}\rightarrow \infty$ provided by the appropriate
choice of the control fields intensity. Maxima of the quantum
efficiency of conversion process reach  $0.996$ for $q_{nf}=100$,
$0.919$ for $q_{nf}=5$ and $0.569$ for $q_{nf}=0$. At greater
magnitudes of $q_{nf}$, the resonance broadening increases, which
leads to the requirement of longer nonlinear medium. This is seen
from comparison of Figs. \ref{gener5}(a) and \ref{gener5}(b). It
is seen from comparison of Figs. \ref{gener4}(b) and
\ref{gener6}(b) that most of the opportunities for manipulation
appear at moderate detunings $\Omega_{mn}$.
%%%%%%%%%%%%%%%%%%%%%%
\section*{Conclusions}
The technique of quantum control of such processes as ionization,
chemical reactivity through dissociation of molecules and
population transfer, and four-wave mixing is further developed.
Other applications involve discrete and broad-band spectra in
solids. The approach is based on manipulating constructive and
destructive interference of several quantum pathways involving
both discrete energy levels (bound states) and energy continua
(free states). Analytical and numerical solutions of the set of
coupled density-matrix equations and Maxwell equations for
travelling  electromagnetic waves are found for the
continuous-wave regime.

Novel opportunities are shown for suppression or alternatively
enhancement of photophysical processes as well as photochemical
processes related with branching chemical reactions. These
opportunities are associated with the overlap of two laser-induced
continuum structures induced by two control fields, while the
third strong field controls a discrete transition. The expressions
obtained  and  numerical models developed  are used to demonstrate
the feasibility of manipulation of the spectral characteristics of
the absorption and dispersion both for discrete transitions and
the spectral continua in the presence of additional strong laser
radiations. Related opportunities to form laser-induced
transparency and to enhance four-wave-mixing nonlinear-optical
polarizations are investigated as well. It is shown that the
implementation of such nonlinear interference effects makes it
possible to utilize the advantages of multiple resonances and
strong radiations for considerable improvement of the generation
of short-wavelength radiation. Among the important features
discovered is the fact that the constructive or destructive nature
of interference is governed not only by the ratio of the
intensities and by the detunings from the discrete resonances, but
also by the composite Fano parameters for transitions between
high-lying discrete levels over the continuum states. The results
can be generalized to higher-order processes. The matrix elements
of the interaction Hamiltonian should then be replaced with the
corresponding composite multiphoton matrix elements.

%%%%%%%%%%%%%%%%%%%%%%%%%%%%%%%
\section*{Acknowledgments}
This work has been supported in part by the International
Association (INTAS) of the European Community for the promotion of
cooperation with scientists from the New Independent States of the
former Soviet Union (Grant INTAS-99-00019), by the Ministry of
Industry and Science of Russian Federation (The 6-th examination
contest of projects of young scientists of the Russian Academy of
Sciences, Grant \#61) and by the Russian Foundation for Basic
Research (Grant 02-02-16325a).  The authors are grateful to K.
Bergmann for discussions and encouragement over the course of this
work.
%\newpage
%\begin{references}

\end{document}